\documentclass[12pt]{article} 

\usepackage[numbers,sort&compress]{natbib}

\usepackage[left=2.5cm,right=2.5cm,top=2cm,bottom=2cm]{geometry}
\usepackage{fancyhdr}
	\fancypagestyle{plain}{ %
	\fancyhf{} 
	\lhead{\textit{Chemical order and  segregation in \gammaprime (\ce{AlAg2)}}}
	\rhead{J. M. Rosalie, C. Dwyer \& L. Bourgeois}
	\lfoot{\textit{Author accepted manuscript}}
	\rfoot{\thepage}
	}

\usepackage{amsbsy} 
\usepackage{graphicx}
	\graphicspath{{./}{./Figures/}} 
\usepackage{subfigure}
\usepackage[version=3]{mhchem} 
\usepackage{multirow} 
\usepackage{multicol}
\usepackage{booktabs}	
\usepackage{setspace}
\usepackage{xspace}

\newcommand{\gammaprime}{\ensuremath{\gamma^{\prime}}\xspace}
\newcommand{\thetaprime}{\ensuremath{\theta^{\prime}}\xspace}

\usepackage[plainpages=false,
pdftitle={On chemical order and interfacial segregation in gamma' (AlAg2) precipitates},
pdfauthor={Julian~M.~Rosalie, Christian Dwyer and Laure~Bourgeois},
pdfsubject={gamma' precipitates in aluminium-silver alloys},
pdfkeywords={Aluminium alloys, Long-range order, Interface segregation; Stacking faults,  Ledgewise Growth},
colorlinks,
debug,
linkcolor=red,
citecolor=red, 
filecolor=red,
breaklinks=false]{hyperref}
\usepackage{doi} 

\title{On chemical order and interfacial segregation in \gammaprime (\ce{AlAg2}) precipitates} 

\author{Julian~M.~Rosalie$^{a}$
\and Christian Dwyer$^{b,c}$
\and Laure~Bourgeois$^{b,d}$}	
\date{}
\vspace{6pt}

\begin{document}

\maketitle
\noindent
$^a$Structural Materials Unit, National Institute for Materials Science (NIMS), Japan.\\
$^b$Monash Centre for Electron Microscopy, Monash University, 3800, Victoria, Australia.\\
$^c$Ernst Ruska-Centre and Peter Gr\"unberg Institute, Forschungszentrum J\"ulich, J\"ulich D52425, Germany.\\
$^d$Department of Materials Engineering, Monash University, 3800, Victoria, Australia.\\
$^\dagger$Present address, work carried out at $\mathrm{^b}$.

\begin{abstract} 
A detailed study has been carried out on \gammaprime (\ce{AlAg2}) precipitates in Al-Ag and Al-Ag-Cu alloys to reconcile the  conflicting reports on chemical ordering and stacking faults in this phase.  
High angle annular dark field scanning transmission electron microscopy and convergent beam electron diffraction  show no indication of chemical ordering on alternate basal planes of \gammaprime precipitates in alloys aged at 473\,K for 2--23\,h. 
Precipitates were visible as Ag-rich regions with 1--13 fcc$\rightarrow$hcp stacking faults, corresponding to 
\gammaprime platelets with thicknesses ranging from 0.69--6.44\,nm.  There were no systematically absent thicknesses. 
Growth ledges with a riser height equal to the $c$-lattice parameter (0.46\,nm)  were directly observed for the first time. 
Genuine stacking faults within the precipitates were extremely rare and only observed in thicker precipitates.
In precipitates with 1--3 stacking faults  there was also substantial Ag in the surrounding fcc layers of the matrix, indicating that Ag strongly segregated  to the broad, planar precipitate-matrix interfaces. This segregation is responsible for previous reports of stacking faults in \gammaprime precipitates. 
The results indicate that the early stages of \gammaprime precipitate growth  are interfacially controlled.

\end{abstract}


\section{Introduction}

The \gammaprime (\ce{AlAg2}) phase is a well-studied intermetallic phase which precipitates in Al alloys and has been used as a model system to study nucleation \cite{passoja:1971,finkenstadt:2006, RosalieActa2011,voss:1999,Finkenstadt2009}, 
ledge-wise growth \cite{howe:1985a,howe:1987,moore:2000,Sagoe-crentsil1991,AikinPlichta1990} and interface structure and energetics  \cite{howe:1987a,sagoe-crentsil:1987,Ramanujan1992a}.
Despite this extensive body of work, questions remain regarding the existence or extent of chemical ordering and the nature of stacking faults widely reported in this phase \cite{guinier:1942,Guinier1952,NicholsonNutting1961,borchers:1969}.

The \gammaprime phase can be produced by quenching an Ag-containing Al alloy from high temperatures to produce a supersaturated solid solution. 
Heating this solid solution to an intermediate temperature (a process termed ``ageing'') results in the gradual precipitation of plate-shaped \gammaprime precipitates on the \{111\} planes of the Al matrix.

The ground state of the  \gammaprime phase is a hexagonal close-packed structure with composition \ce{AlAg2} and short-range order (SRO) on the basal planes  \cite{neumann:1966,zarkevich:2002}.
Experimental reports indicate that in practice the \gammaprime phase departs considerably from stoichiometry.
An atom probe field ion microscopy study by Osamura \textit{et. al.} in Al-5.72at.\%Ag aged at 436K reported an Ag concentration  of 33.3$\pm1.5$at.\% \cite{Osamura1986} and more recent studies using energy dispersive X-ray analysis in Al-22at.\%Ag found an average composition of 42at.\%Ag  \cite{moore:2000}.

X-ray studies  to date have also failed to find evidence of short-range order  \cite{neumann:1966,YuGammaprime2004} within the basal plane. 
Monte Carlo simulation studies have suggested an order-disorder transition temperature in the range of 45--100\,K  \cite{YuGammaprime2004,neumann:1966}.
Such temperatures are well below the ageing temperatures required to form \gammaprime precipitates in a finite time scale and this might explain the absence of SRO.  

There have also been conflicting reports of chemical ordering on alternate basal planes (i.e. long-range order) in \gammaprime precipitates.  
Although early X-ray studies proposed a ground state with alternate layers of \gammaprime precipitates having compositions of \ce{Al2Ag} and \ce{AlAg2},  these studies found compositional variations between alternate planes of at most 1.5at\% \cite{YuGammaprime2004,neumann:1966}.
 A structure consisting of layers with compositions Ag and \ce{Al2Ag} has also been suggested based on high resolution transmission electron microscopy (HRTEM)  studies \cite{howe:1985a,howe:1987} that detected high and low contrast layers  \cite{howe:1987a} on alternate basal planes.

Clarifying the existence and/or extent of ordering in \gammaprime is important for models of the structure and growth of this phase.
Recent density functional theory (DFT) simulations have modelled the structure as having well-defined order on alternate basal planes \cite{finkenstadt:2006,Finkenstadt2009,Finkenstadt2010} and may require revision to represent the structure at non-zero temperatures\footnote{DFT calculations are carried out at 0\,K.}  if the phase is disordered.

A further area of uncertainty regards stacking faults which have been widely reported in \gammaprime precipitates.
The earliest reports of such defects were by Guinier \cite{guinier:1942,Guinier1952}  who reported stacking faults with an average spacing of only 10\,\AA~based on  small angle X-ray scattering.  
Similar results were obtained via electron diffraction by Nicholson and Nutting \cite{NicholsonNutting1961} who also used the width of selected area diffraction peaks to determine the fault spacing as a function of precipitate thickness. 
These authors reported similar fault spacings to Guinier \cite{guinier:1942,Guinier1952} and Borchers \cite{borchers:1969} for thin precipitates, and an increase in the spacing as the precipitate thickness  increased to 80\,\AA.
Despite this, HRTEM studies on thin (1-3 unit cell) precipitates did not show abundant stacking faults within  \gammaprime precipitates  \cite{RosalieActa2011}. 
The discrepancies between diffraction and imaging studies have not been resolved. 

The present work sets out to re-examine the structure of \gammaprime precipitates using high-resolution scanning transmission electron microscopy. 
The absence of complicated phase contrast in high angle annular dark field scanning transmission electron microscopy  (HAADF-STEM), together with the insensitivity of this technique to strain contrast make it ideal for detecting chemical  ordering. 
These capabilities are also well-suited for examining the structure of the precipitate in order to detect and characterise any stacking faults. 

\section{Experimental details}

Aluminium alloys with compositions Al-1.68at.\%Ag (Al-6.4wt.\%Ag) and Al-0.90at.\%Cu-0.90at.\%Ag prepared from pure elements were used in this study.
Billets of each alloy were cast in air at 973\,K and poured into graphite-coated steel molds. 
The compositions and impurity levels  were measured by inductively-coupled plasma atomic emission spectrometry.  
Impurity levels were low, with Fe present at 0.025\,wt.\% and Si at 0.01 wt.\%.
The Ni content was 0.01 wt.\% with other elements (Cu, Zr, Ti, Mn, Mg) present at levels of  $<0.005$ wt.\%.

Billets were homogenised  (798$^\circ$C, 168\,h)   then hot-rolled (to 2\,mm thickness) and cold-rolled to produce 0.5\,mm thickness sheet. 
Discs (3\,mm diameter) punched from the sheets were  solution-treated (525$^o$C, 0.5\,h) in a nitrate/nitrite salt pot and then  quenched to room temperature in water. 

Solution-treated samples of Al-Ag alloy were aged in an oil bath 473\,K for 2--23\,h and quenched into water. 
Al-Ag-Cu alloys were aged for 2--4\,h under the same conditions. 
The aged discs were  mechanically thinned and then twin-jet polished to perforation using a nitric acid/methanol solution ($\sim$13\,V, 253\,K in  33\% \(\mathrm{HNO_3}\)$/$67\%\(\mathrm{CH_3OH}\) v/v).
Discs were plasma-cleaned  immediately prior to examination.
	
Foils were examined using a FEI Titan$^3$ 80-300 microscope operating at 300\,kV.
High angle annular dark field scanning transmission electron microscope (HAADF-STEM) images were obtained using a  convergence semi-angle of 15\,mrad, providing a spatial resolution of $\approx 1.2$\,\AA, using an inner collection angle of 40\,mrad.  

The HAADF-STEM images were compared with images calculated by a multi-slice method using a frozen phonon approach to incorporate thermal diffuse scattering \cite{DwyerStem2010}.
The simulations used the \gammaprime structure proposed by Neumann, with a slice thickness of 1.4 \AA~ and a maximum thickness of 700 \AA.
Atomic sites in the simulated structures were randomly occupied within each layer and alternate layers had compositions of \ce{Ag2Al} and \ce{AgAl2}. 
These were compared with simulations in which all layers had identical compositions.

Image analysis was carried out using ImageJ software (version 3.6322U2011).
Simulated images were scaled linearly to the same dynamic range as the HAADF-STEM images. 
The contrast in simulated images was then adjusted to match the contrast of the experimental images by modifying the gamma correction value ($\Gamma$). 
The adjusted intensity ($I^\prime$) was given by 
\( I^\prime = I^{1/\Gamma} \).
Except where otherwise stated, all experimental and simulated images presented were adjusted for brightness and contrast in this manner,  with no other manipulations. 

Convergent beam electron diffraction (CBED) was performed in STEM mode using a JEOL 2100F instrument operating at 200 kV, with a 0.5\,nm probe. CBED maps were obtained using the Diffraction Imaging plug-in of Gatan DigitalMicrograph software on foils aged for 23\,h at 473\,K. 

\section{Results}

\subsection{HAADF-STEM imaging}

The \gammaprime precipitates are readily visible in HAADF-STEM images, due to the strong atomic contrast of Ag compared to Al.
\gammaprime precipitates in the Al-Ag-Cu alloys form in complex assemblies along with \thetaprime precipitates. 
For ageing times of 2--4\,h, the \gammaprime precipitates are identical in appearance to those in the binary alloy.
However, the presence of \thetaprime precipitates disrupts the growth of the \gammaprime phase \cite{RosalieThetaPrimesilver2012} and for ageing times of $\ge$4\,h, few \gammaprime precipitates are observed. 

Throughout this work the local structure is described in terms of the stacking sequence for a given close-packed plane; hcp for the  \gammaprime precipitate and fcc for the Al matrix. In addition, the precipitate-matrix interface is coherent and the matrix layers in contact with the precipitate are denoted ``fcc/hcp'' to highlight the fact that such layers are common to both the precipitate and matrix.
In keeping with our previous work  \cite{RosalieActa2011} the thickness of the \gammaprime precipitates is described in terms of the number of fcc$\rightarrow$hcp stacking faults. 
This allows a clear distinction to be drawn between the thickness of the precipitate itself and the size of the region which is enriched in Ag. This distinction is central to the discussion of solute segregation in the following sections.

In foils of Al-Ag and Al-Ag-Cu aged for 2--4\,h at 473\,K, HAADF-STEM images show that the majority of precipitates 
contain either 1 or 2 fcc$\rightarrow$hcp stacking faults (hereafter denoted as 1--2$c$(\gammaprime)).
In foils aged for 23\,h, precipitates containing up to 13 stacking faults (i.e. 13$c(\gammaprime)$) are observed.
The precipitate thickness are measured for all ageing times for both alloys and it is noted that all integer values of thickness (number of stacking faults) are present i.e. there are no systematically missing thicknesses.

\begin{figure}[htbp]
	\begin{center}
		\hfill
		\subfigure[$1c$ (Al-Ag-Cu, 4\,h)\label{1c}]{\includegraphics[width=0.35\textwidth]{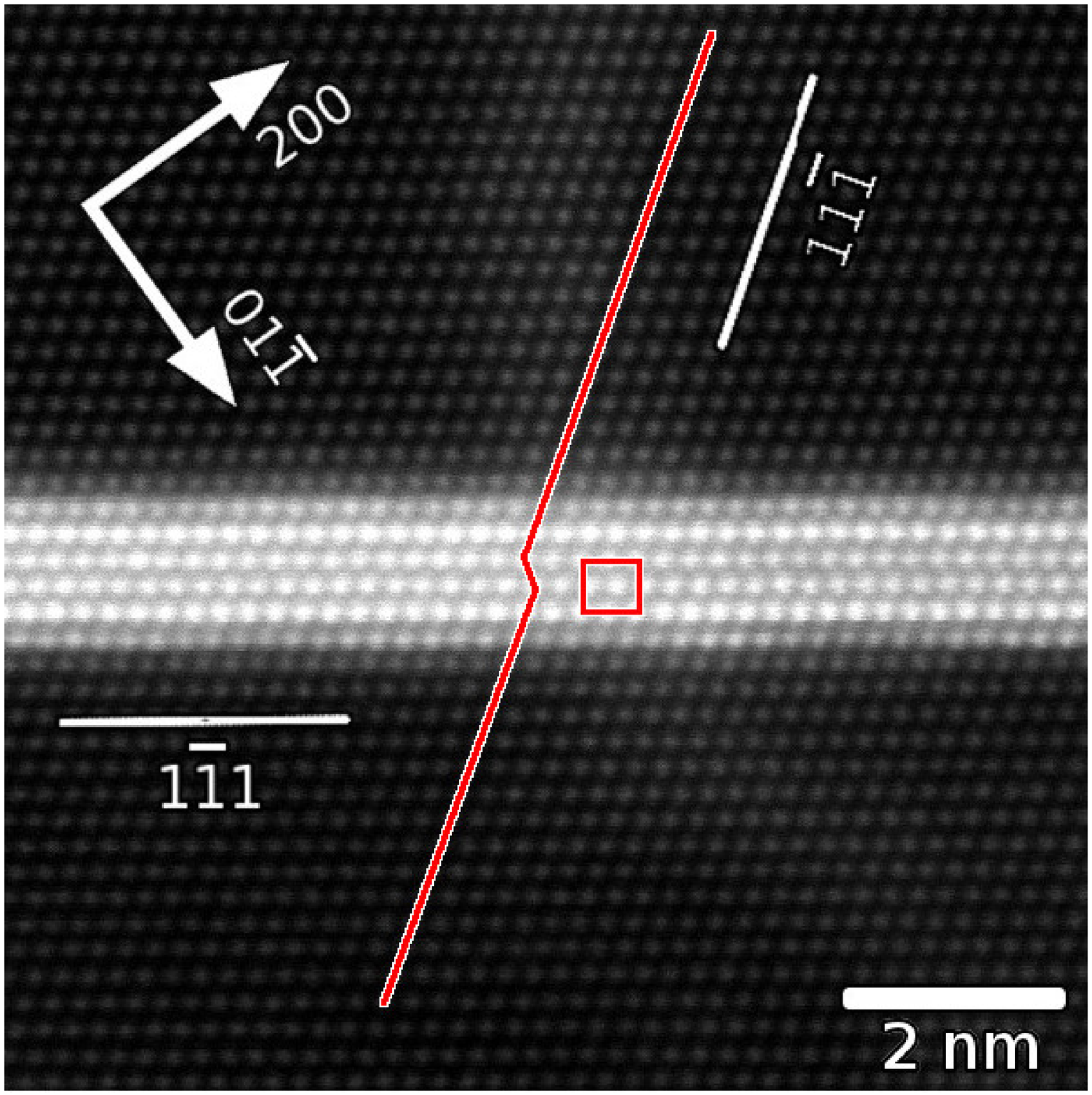}}
		\hfill
		\subfigure[$2c$ (Al-Ag, 4\,h) \label{2c}]{\includegraphics[width=0.35\textwidth]{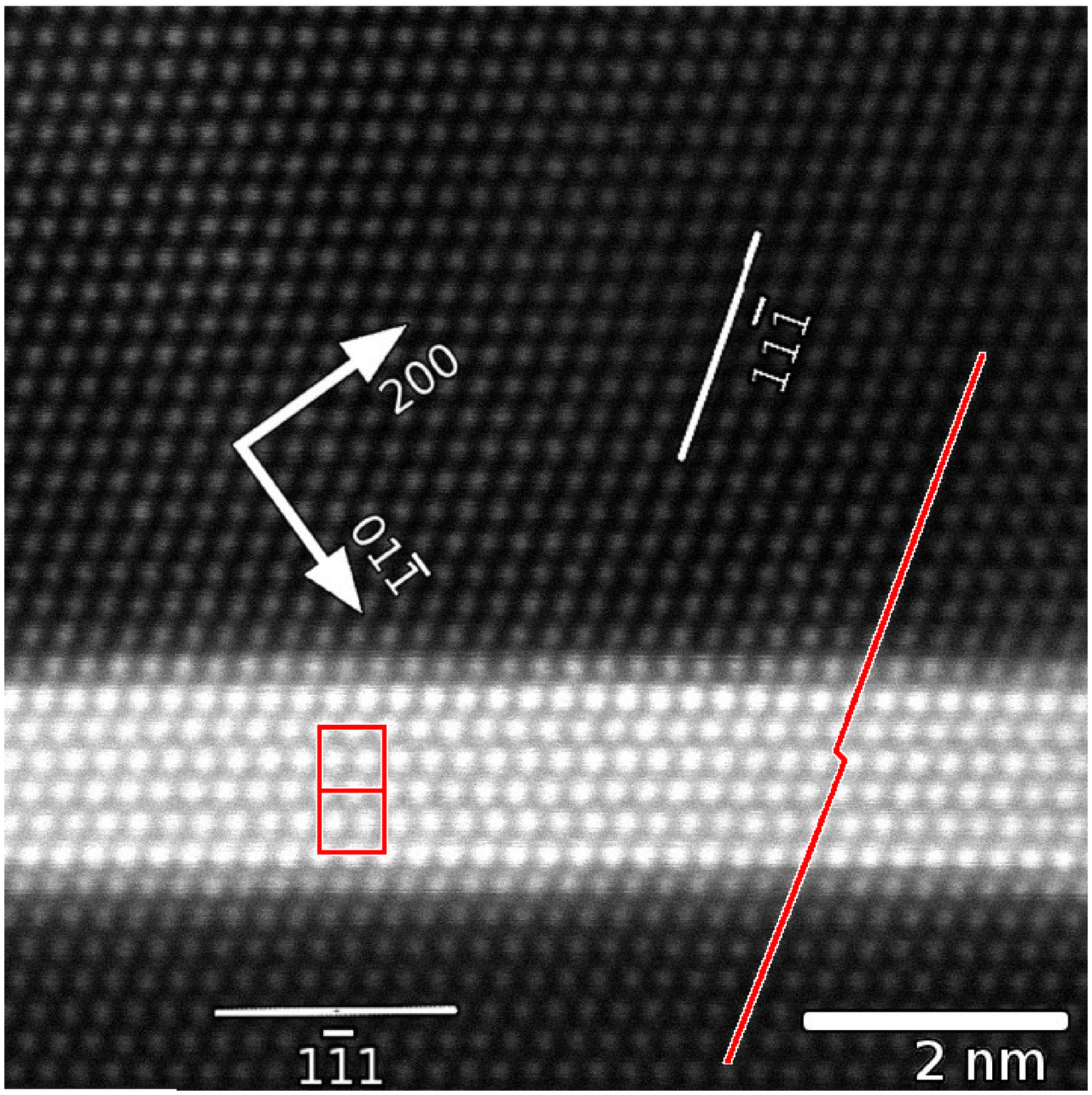}}
		\hfill\

		\hfill
		\subfigure[$3c$ (Al-Ag-Cu, 4\,h) \label{3c}]{\includegraphics[width=0.35\textwidth]{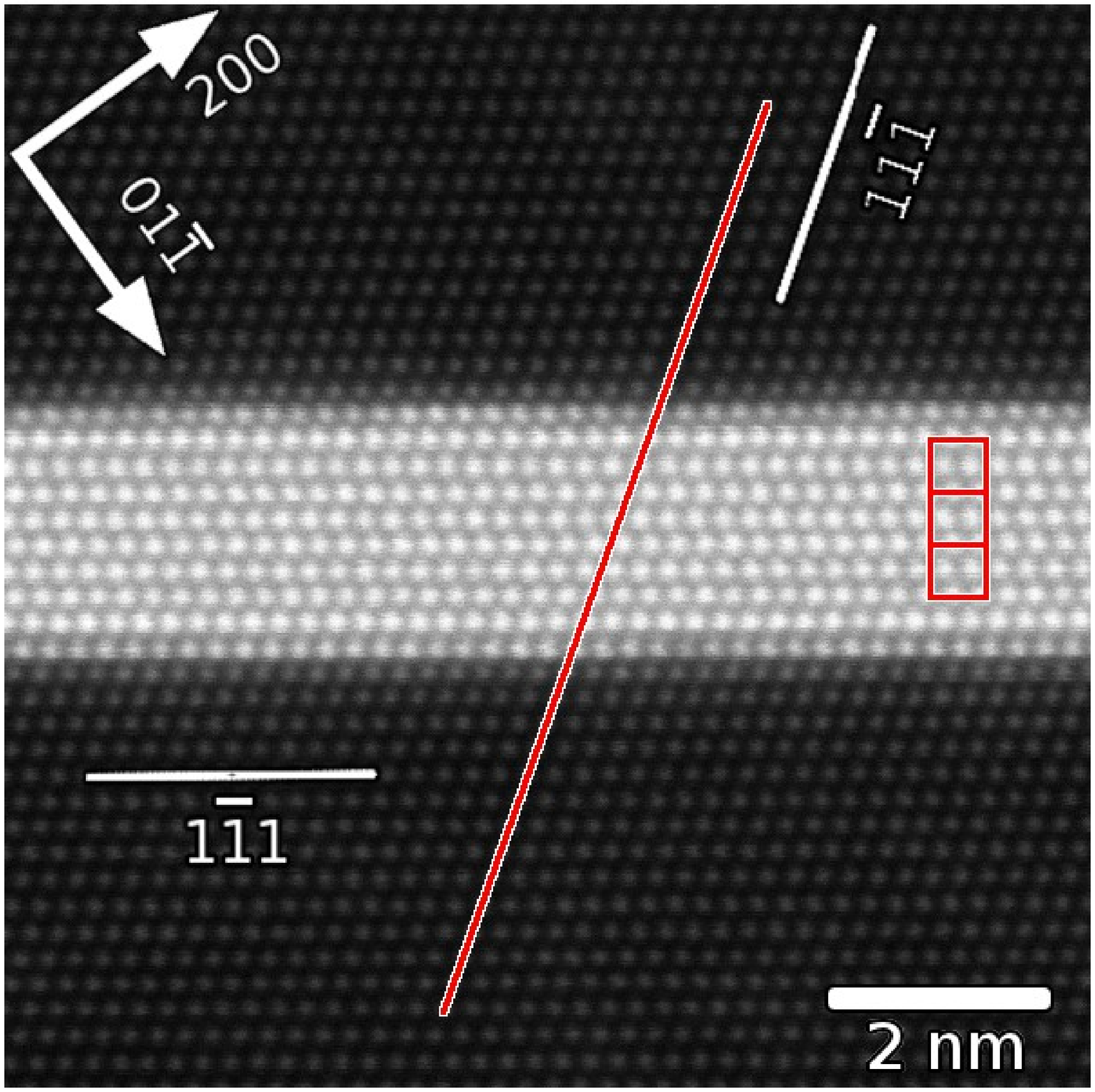}}
		\hfill
		\subfigure[$4c$ 
\newline Al-Ag (2\,h) \label{4c}]{\includegraphics[width=0.35\textwidth]{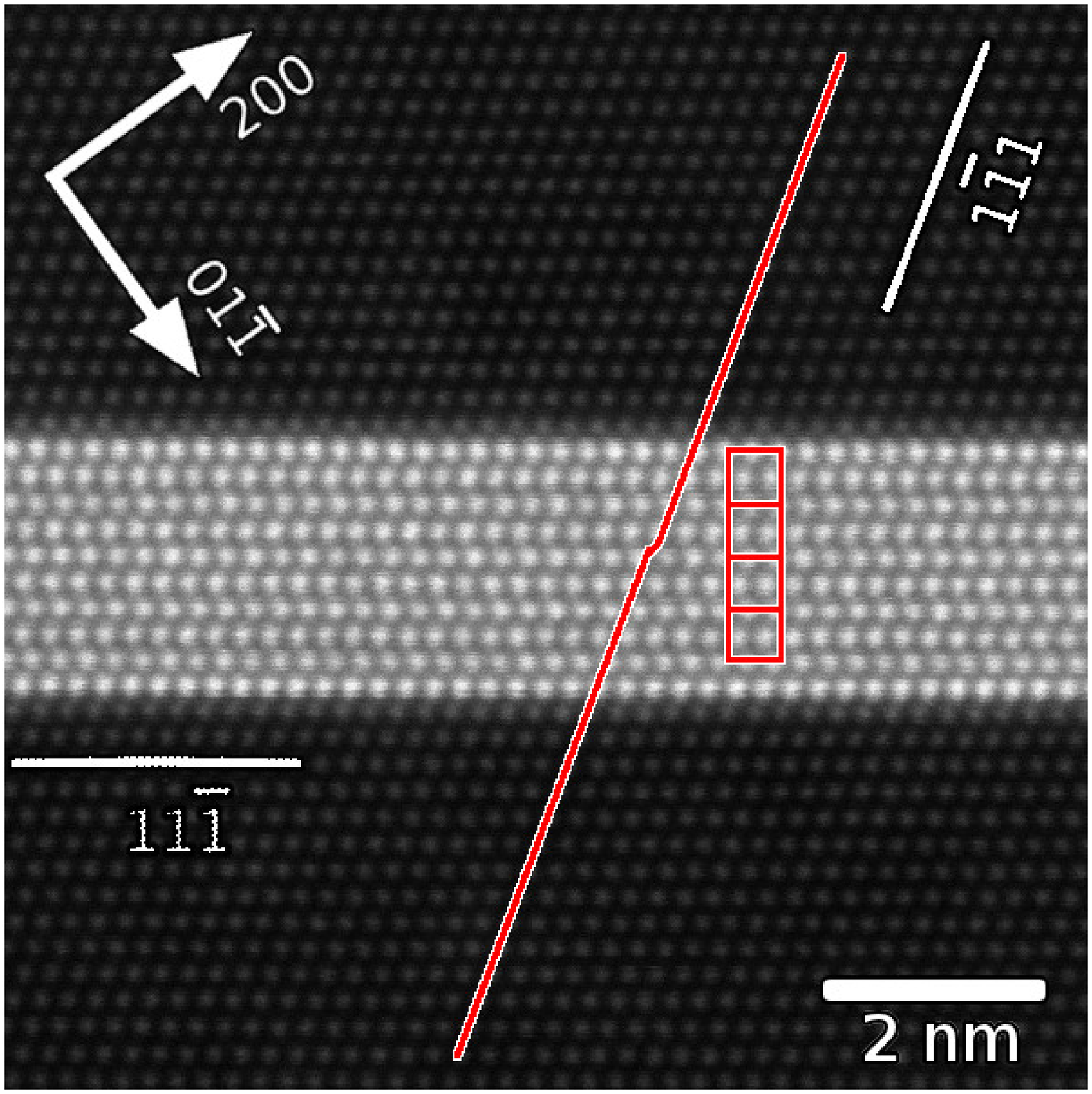}}
		\hfill\

		\hfill
		\subfigure[$7c$ (Al-Ag-Cu, 4\,h) \label{7c}]{\includegraphics[width=0.35\textwidth]{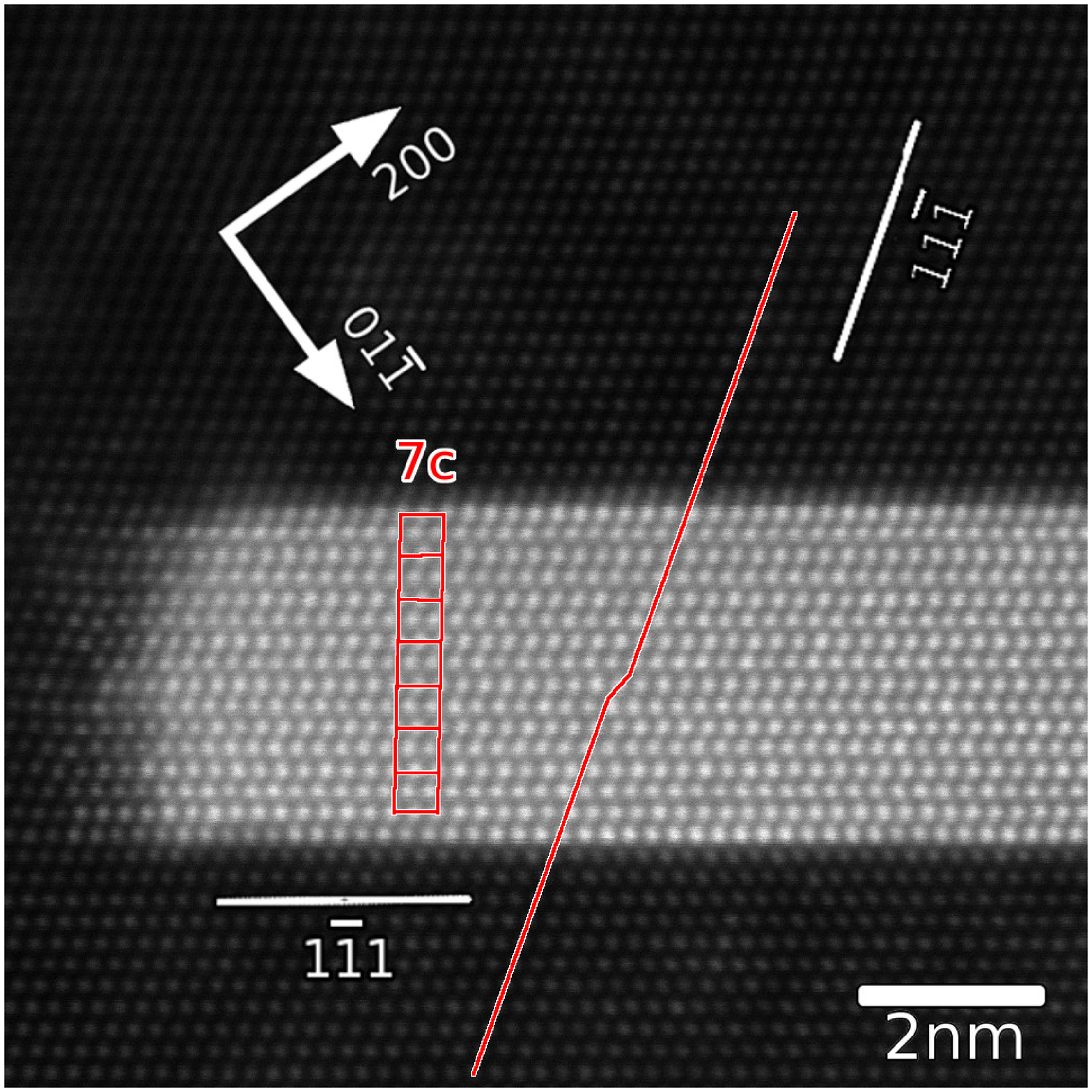}}
		\hfill
		\subfigure[$12c$ (Al-Ag, 23\,h) \label{12c}]{\includegraphics[width=0.35\textwidth]{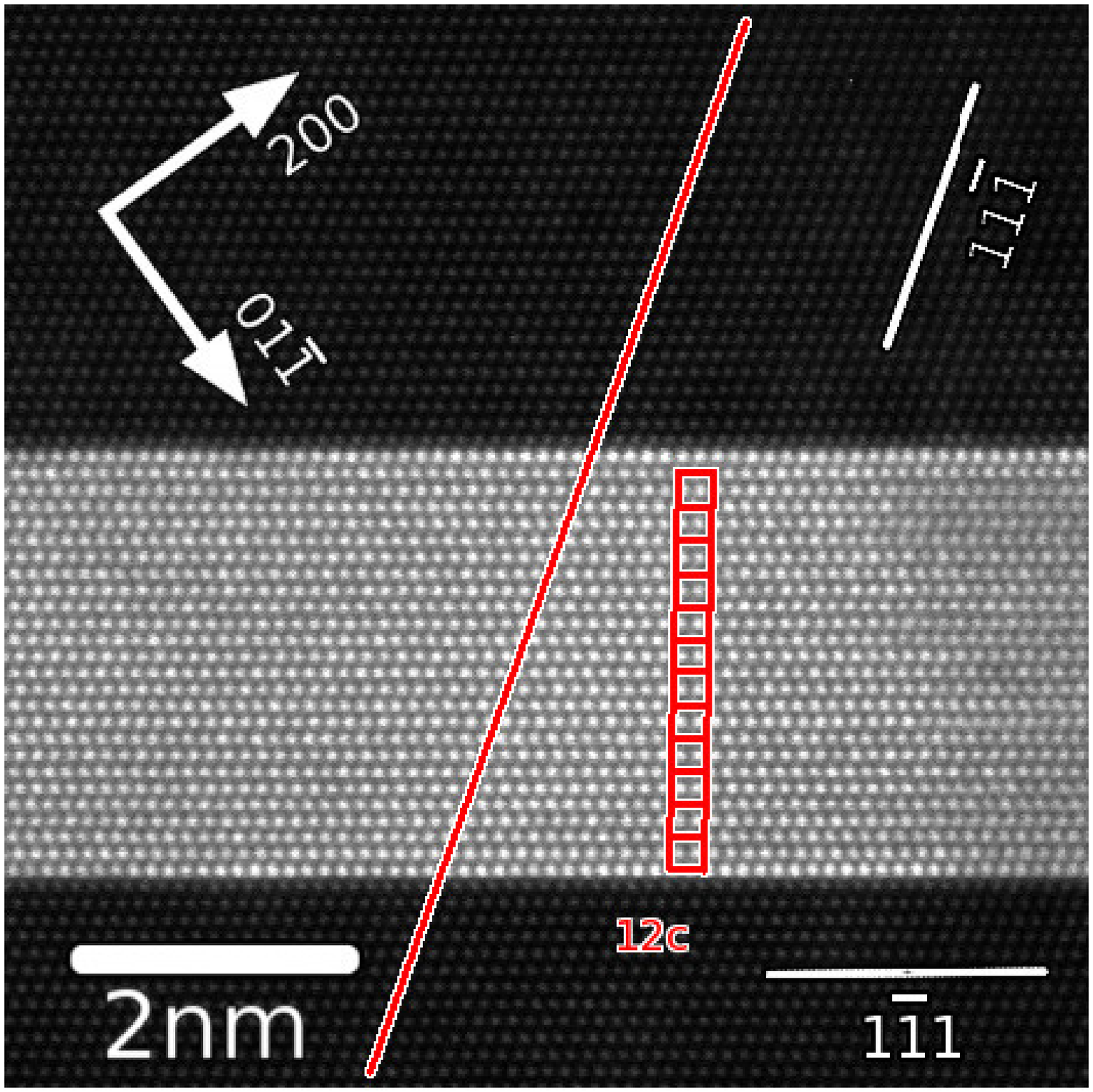}}
		\hfill\
		\caption{HAADF-STEM images of \gammaprime precipitates of with different numbers of stacking faults. The alloy and ageing time are given in parentheses.
The overlays indicate the hcp region. Lines drawn across the precipitate indicate whether the matrix is incommensurate  (a, b, d, e) or commensurate (c, f). 
  \label{fig-haadf-series}}
	\end{center}
\end{figure}

Figure~\ref{fig-haadf-series} shows representative HAADF-STEM images of precipitates with thicknesses ranging from 1--12$c(\gammaprime)$.\footnote{To assist the reader, the micrograph in Fig. 1(d) has been flipped so as to present all precipitates in a consistent orientation.}
For precipitate thickness of 1--3$c(\gammaprime)$ Ag  appears to be heavily enriched outside the hcp region  (Figures~\ref{1c}--\ref{3c}).
This can be most clearly seen for the precipitate in Figure~\ref{1c}. 
A single stacking hcp$\rightarrow$fcc stacking fault produces only 2 hcp close-packed layers and 2 interface fcc/hcp layers, yet there is strong atomic contrast due to Ag across 6 close-packed layers.
This indicates Ag segregation to the matrix-precipitate interface and is  discussed in further detail in Section~\ref{sec-segregation}. 

For precipitates of  3$c(\gammaprime)$ and 12$c(\gammaprime)$ thickness (Figures~\ref{3c},\ref{12c}) the matrix is commensurate and atomic columns on either face of the precipitate are well aligned.
If the image is viewed at a glancing angle, it can be seen that lines drawn along atomic columns on one side of the image fall along atomic columns on the opposite face of the precipitate.
For precipitates with thicknesses of 1, 2, 4 and 7 $c(\gammaprime)$ the matrix is incommensurate and this is not the case.
Since the planar interfaces remain coherent with the matrix, this indicates that the lateral displacements associated each unit cell of the \gammaprime phase are self-accommodated for $3 n\, c(\gammaprime)$ \cite{RosalieActa2011} where $n$ is the number of stacking faults.

\begin{figure}[htbp]
	\begin{center}
		\hfill	
		\subfigure[5--6$c$ (Al-Ag, 4\,h) \label{5-6c}]{\includegraphics[width=0.35\textwidth]{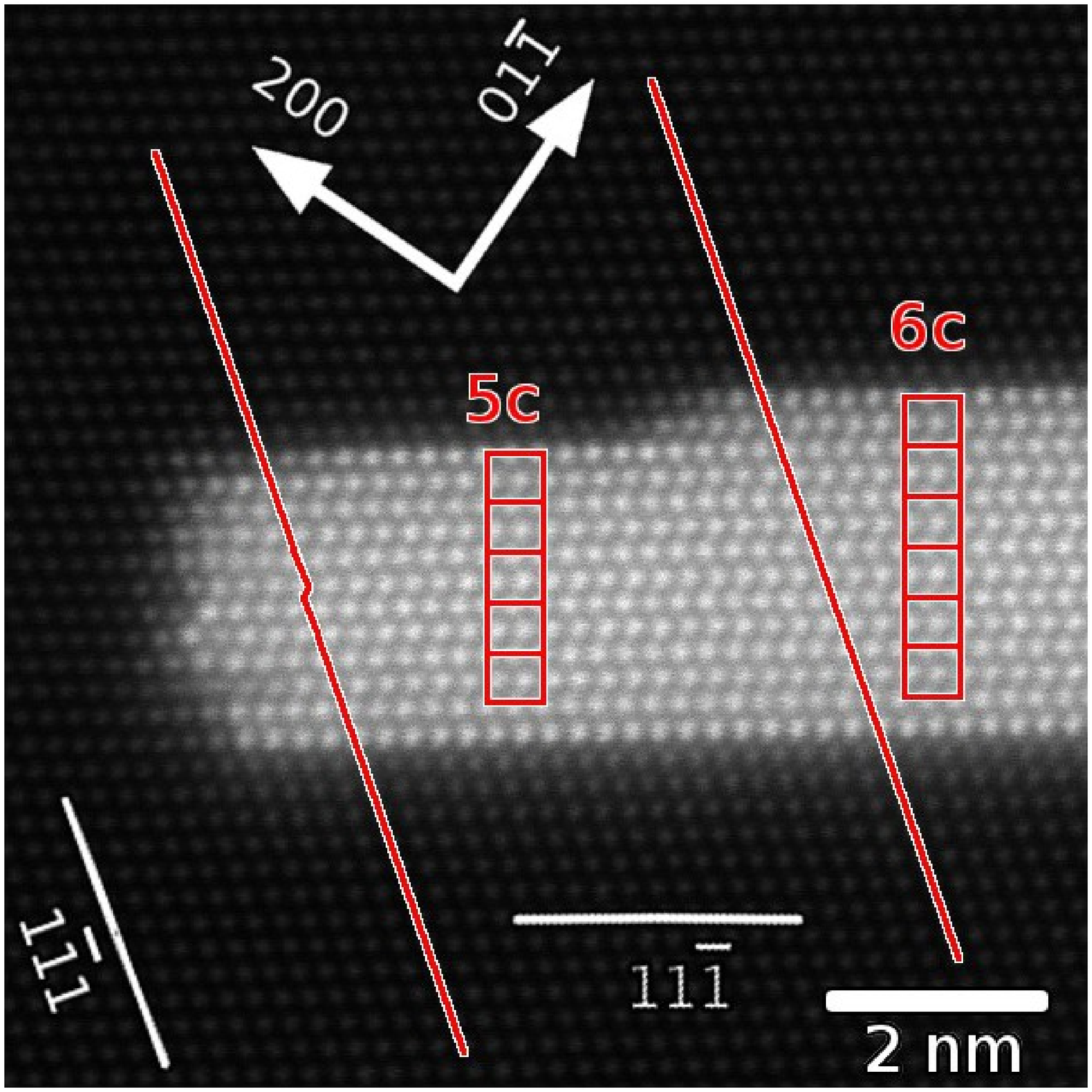}}	
		\hfill
		\subfigure[12--13$c$ (Al-Ag, 23\,h) \label{12-13c}]{\includegraphics[width=0.35\textwidth]{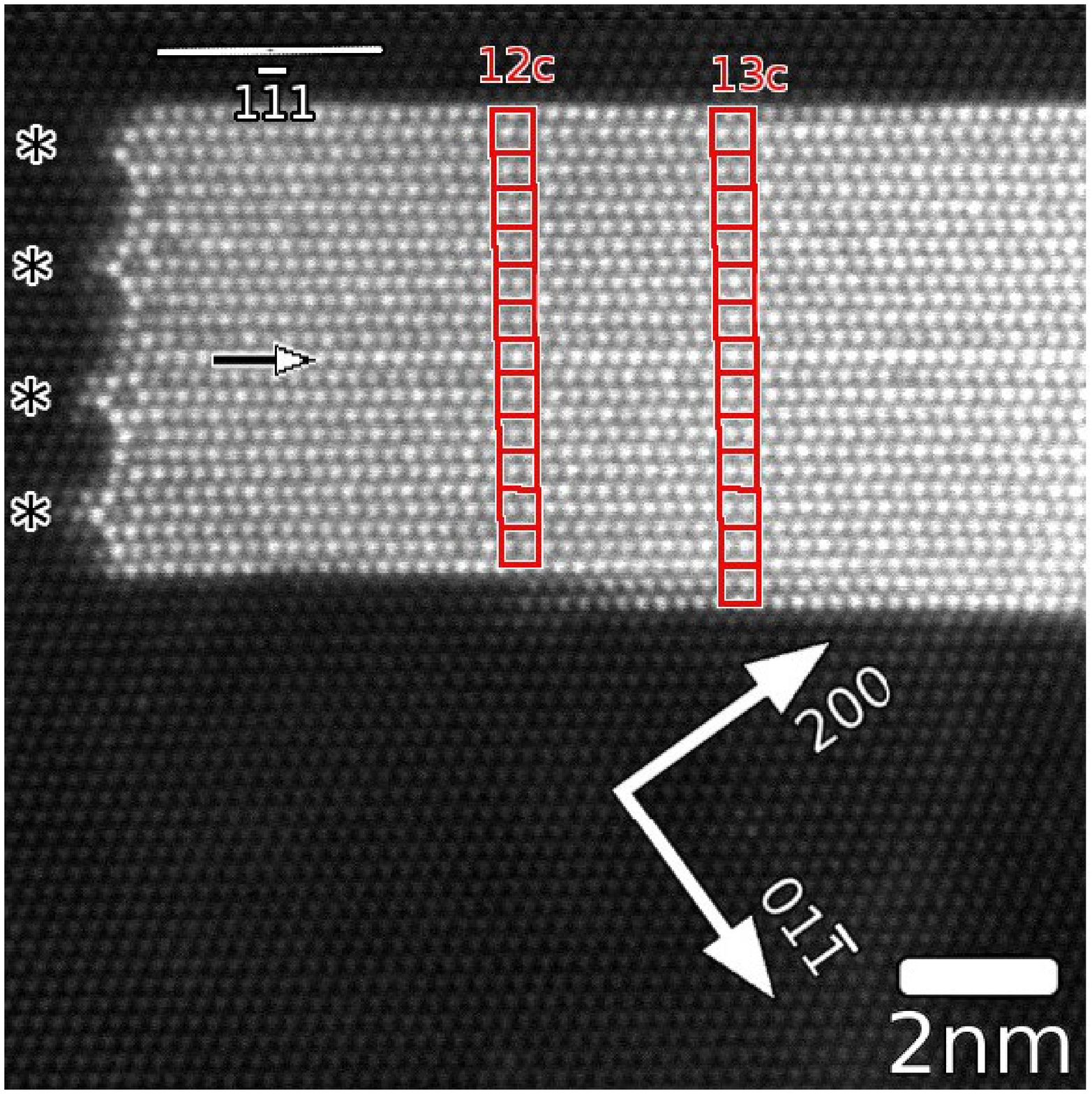}}
		\hfill\
		\caption{HAADF-STEM images of growth ledges on \gammaprime precipitates. 
The overlays and associated labels indicate the  thickness of \gammaprime platelets in terms of stacking faults.   The alloy and ageing time are given in parentheses.
The ledge height is  1c$(\gammaprime)$ for precipitates of thickness (a) 5c$(\gammaprime)$ and (b)  12c$(\gammaprime)$.
Lines have been drawn across the precipitate in (a) and show that the matrix is incommensurate for thicknesses of 5$c$ and commensurate for 6$c$. 
The asterisks in (b) indicate Shockley partial dislocations at the semi-coherent edges of the precipitate.
\label{fig-haadf-series1}}
	\end{center}
\end{figure}

Growth ledges are uncommon, but those which are observed invariably have a ledge riser height equivalent to one unit cell of \gammaprime.  
Figure~\ref{fig-haadf-series1} shows examples of ledges on precipitates with thicknesses of (a) 5$c(\gammaprime)$ and (b) 12 $c(\gammaprime)$.
In all cases where ledges were observed the precipitate thickness was greater at the centre than the edge, suggesting that ledge nucleation occurs near the centre of the precipitate. 
Lines drawn along atomic columns indicate that the precipitate in Figure~\ref{5-6c} is commensurate at the centre (i.e. thickness=$6c(\gammaprime)$) and incommensurate close to the edge (where thickness=$5c(\gammaprime)$).  

The semi-coherent edges of the precipitate in Figure~\ref{12-13c} have a saw-tooth pattern with a total of four ridges present (marked by asterisks in the figure).  
The hexagonal features present at each ridge probably indicate the core of Shockley partial dislocations, one of which should be parallel to the electron beam direction in every third unit cell. 
The spacing between the uppermost ridges and the adjacent ridges is $\sim$1.4\,nm, corresponding to 3$c$(\gammaprime).
A similar separation exists between the lower ridge and the adjacent ridge. 
However, the spacing between the central pair of ridges is 4$c$(\gammaprime).
This is related to a change in structure occurring at the central plane of the precipitate in Figure~\ref{12-13c} (indicated by an arrow in the figure) which shows noticeably stronger contrast than the surrounding planes.
This plane (indicated by an arrow in the figure) is the site of a single hcp$\rightarrow$fcc stacking fault, which is described in greater detail in Section~\ref{sec-stacking-faults}.

\begin{figure}[htbp]
	\begin{center}
		\includegraphics[width=8cm]{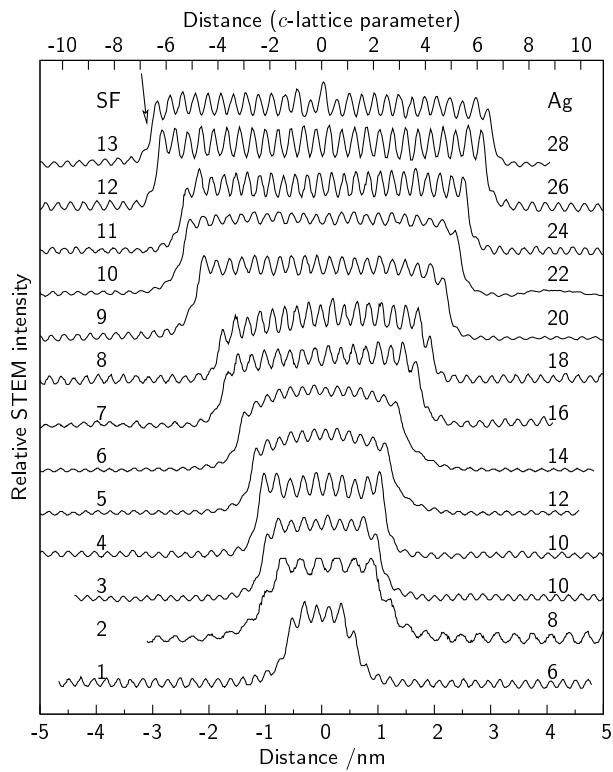}
		\caption{HAADF-STEM intensity profiles for \gammaprime precipitates with  1--13 stacking faults 
i.e. $n$=1 to 13$c(\gammaprime)$. 
The number of stacking faults (``SF'', left) and Ag-rich layers (``Ag'', right) are included in the figure. 
There is no indication of alternating high and low intensity layers that would indicate chemical ordering on alternate basal planes. \label{fig-line-profiles}}
	\end{center}
\end{figure}

Analysis of intensity profiles of the HAADF-STEM images shows no evidence of long-range order within \gammaprime precipitates for ageing times up to at least 23\,h at 473\,K.
This can be seen most clearly by examining line profiles of the HAADF-STEM intensity across the precipitate, as is done in Figure~\ref{fig-line-profiles}.
Line profiles were drawn normal to the precipitate matrix interface, using an integration width of 0.8\,nm. 
The figure shows representative intensity profiles for precipitates with thicknesses ranging for 1 to 13 $c(\gammaprime)$.
Each profile has been linearly scaled from 0 to 1 in order to show the relative HAADF intensity across the precipitate. 
There are slight fluctuations between the intensity of adjacent layers; however, there are no systematic alternations in the intensity that would indicate layers of higher or lower Ag content. 
The intensity profiles in Figure~\ref{fig-line-profiles} also show the greater-than-expected thickness of the Ag-enriched region, as  noted in Figure~\ref{fig-haadf-series}.

\subsection{Stacking faults in \gammaprime precipitates \label{sec-stacking-faults}}

\begin{figure}[htbp]
	\begin{center}
		\hfill
		\subfigure[\gammaprime-Al interface showing change of stacking.\label{fig-stacking-1}]{\includegraphics[width=4.2cm]{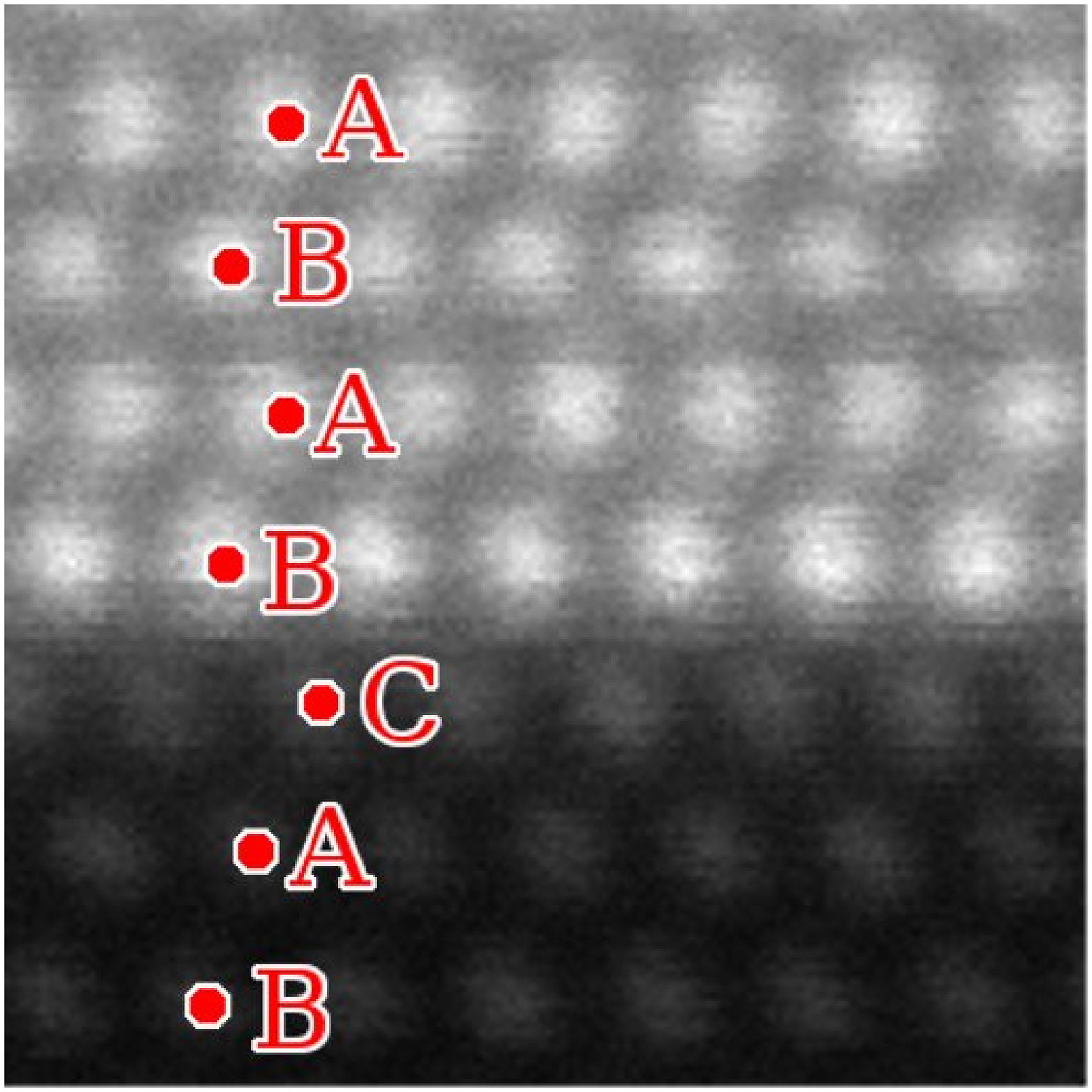}}
		\hfill
		\subfigure[\gammaprime-Al matrix with surface stacking fault in \gammaprime phase.\label{fig-stacking-2}]{\includegraphics[width=4.2cm]{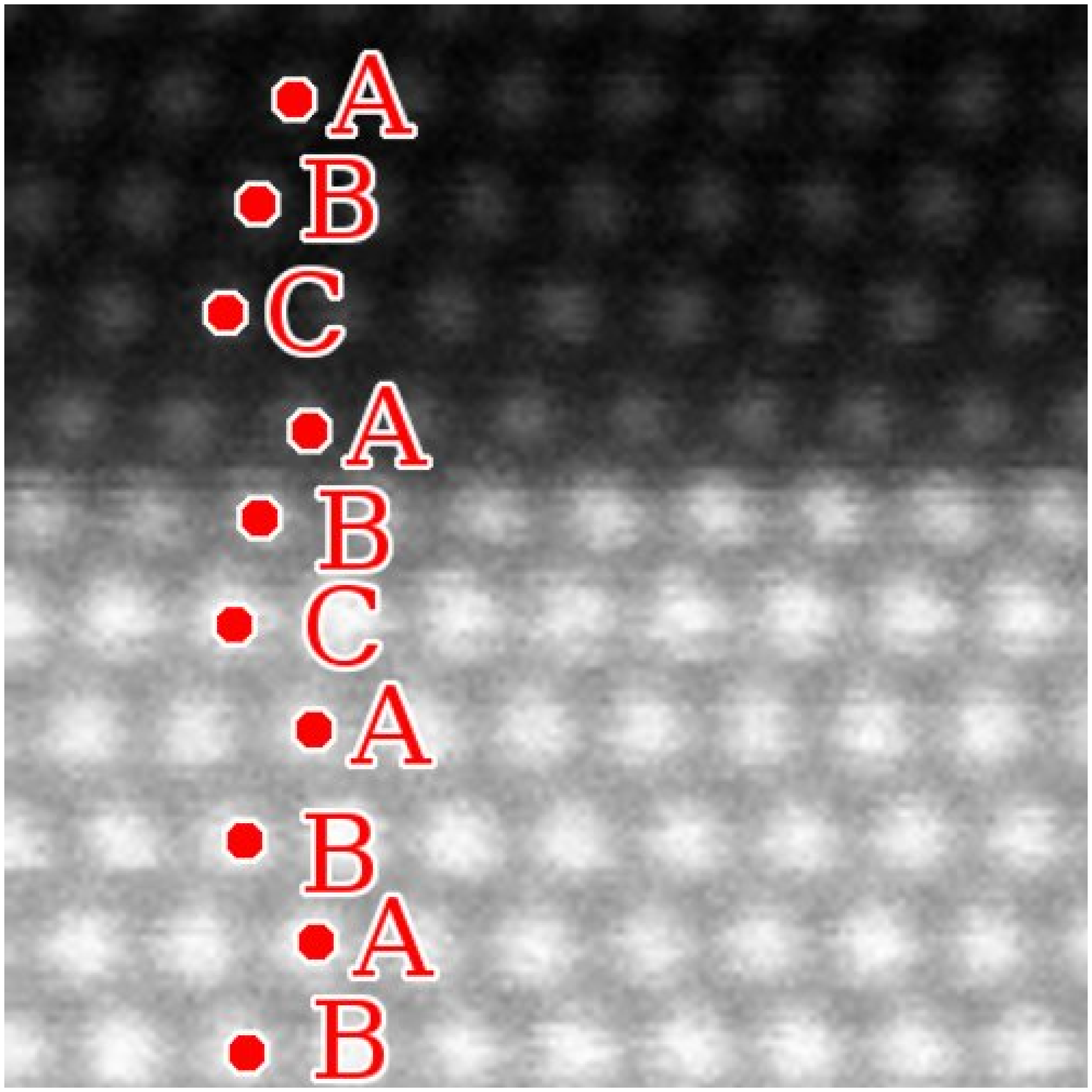}}
		\hfill
		\subfigure[\gammaprime precipitate with internal stacking fault.\label{fig-stacking-3}]{\includegraphics[width=4.2cm]{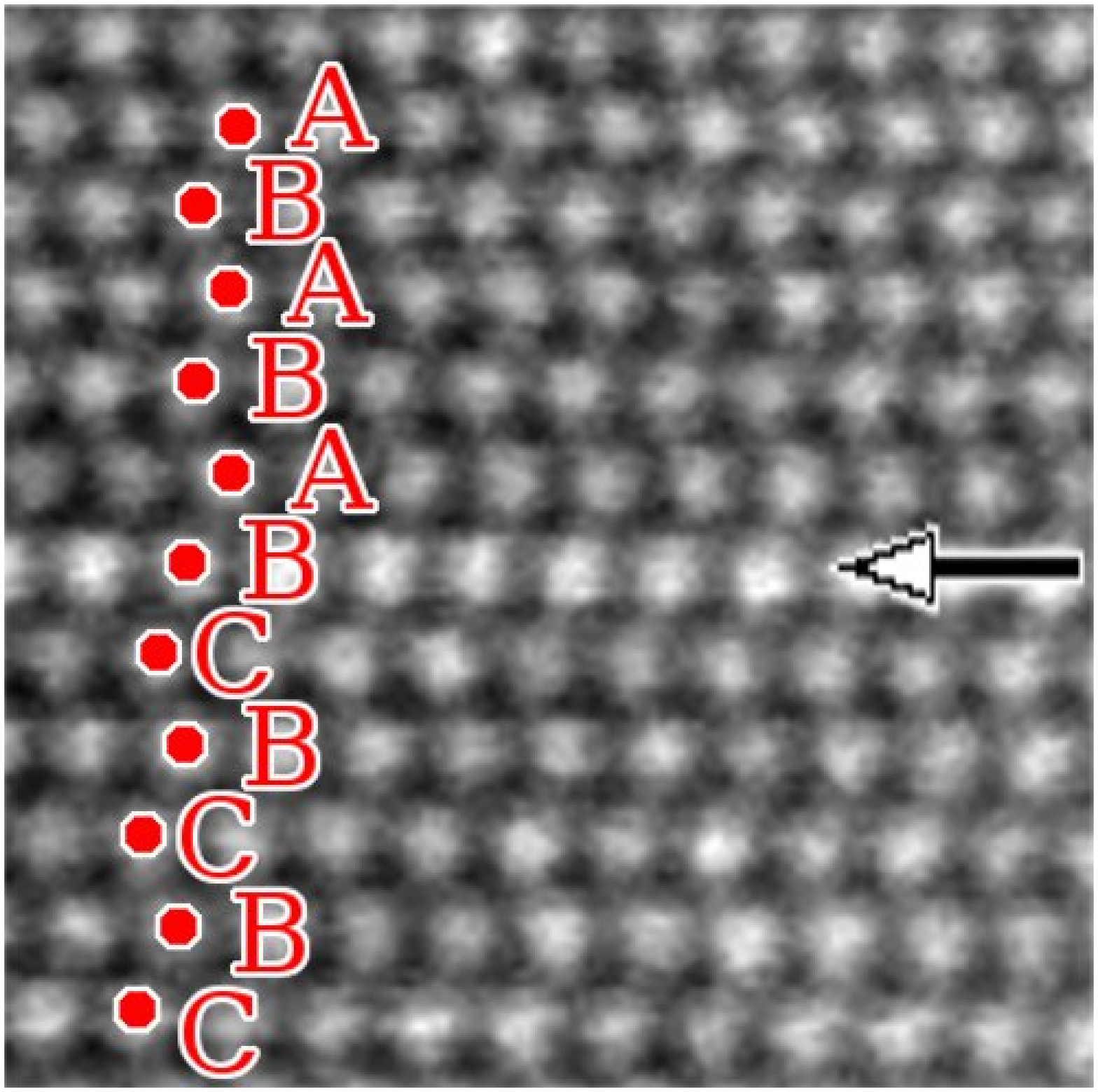}}
		\hfill\
		\caption{Changes in the stacking sequence at the \gammaprime-Al interface and within \gammaprime precipitates.
(a) shows a \gammaprime-Al interface with the change in stacking sequence from fcc $ABC$ stacking to hcp $AB$ at the boundary of the Ag-rich region.
(b) shows the \gammaprime-Al interface of a precipitate in which the outer Ag-rich planes retain the fcc stacking sequence.
(c) shows the central portion of a  \gammaprime precipitate with a stacking fault on the plane indicated by the arrow.
\label{fig-stacking}}
	\end{center}
\end{figure}

Changes in the stacking sequence are readily identifiable in the HAADF-STEM images.
Representative micrographs illustrating this are provided in Figure~\ref{fig-stacking}.
For precipitates of thickness greater than 3--5$c(\gammaprime)$ there is a clear change in stacking sequence coinciding with the Al-\gammaprime compositional interface (i.e. where the image contrast, and hence composition, changes most rapidly). 
Figure~\ref{fig-stacking-1} shows the matrix-precipitate interface of a \gammaprime precipitate of thickness 7 $c(\gammaprime)$. 
The change in stacking from hcp ($AB$) in the precipitate to  fcc in the matrix ($ABC$) coincides precisely with the compositional change evident in the HAADF-STEM contrast. 
The outermost Ag-rich $B$ layer is common to both crystal structures (such layers are described as ``fcc/hcp'' to denote this).

In precipitates with  1--3 stacking faults  the compositional interface is not coincident with the change in stacking sequence and there are Ag-rich atomic planes outside the hcp region.
Figure~\ref{fig-stacking-2} shows an enlarged image of the interface region of the $3c(\gammaprime)$ precipitate from Figure~\ref{3c}.
The interior of the precipitate (in the lower region of the figure) has the hcp structure with $AB$ stacking; however the two uppermost Ag-rich layers have fcc ($ABC$) stacking. 

Deviations from the hcp stacking sequence within the precipitates are extremely rare and only observed for thicker precipitates.
Figure~\ref{fig-stacking-3} shows one example taken from the central region of the precipitate shown in Figure~\ref{12-13c}.
As noted previously, the mid-plane of the precipitate (indicated by an arrow) is somewhat stronger in atomic contrast than the surrounding layers. 
Inspection of the micrograph reveals that this plane is also the site of a stacking fault, with the stacking sequence changing from $A$--$B$ above the fault to $B$--$C$ below this plane.

\subsection{Chemical ordering and silver segregation\label{sec-segregation}}
Multi-slice image simulations were performed in order to determine firstly, whether long-range order of Ag and Al should be discernible under the present experimental conditions, and secondly whether the bright columns in fcc regions adjacent to \gammaprime are due to Ag segregation or electron scattering effects.

Of the two proposed structures for the \gammaprime precipitate, 
Neumann's model  \cite{neumann:1966} which is comprised of \ce{Ag2Al} and \ce{AgAl2} layers has less atomic number difference between alternate layers than the structure consisting of \ce{Ag2Al} and \ce{Al} layers proposed by Howe \cite{howe:1987}.
The average atomic numbers for the layers in  the Neumann structure are 35.7 and 24.3 (a ratio of 1.48:1), whereas for the Howe structure the average atomic numbers are 35.7 and 13 (a ratio of 2.74).
This will result in less atomic contrast between layers for the Neumann structure and should provide a more rigorous test of the sensitivity of HAADF-STEM to chemical order in \gammaprime precipitates.

\begin{figure}[htbp]
\begin{center}
\hfill
\subfigure[Ordered \ce{AlAg2} (Neumann model) \label{fig-simulation-ordered}]{\includegraphics[width=0.25\textwidth,angle=90]{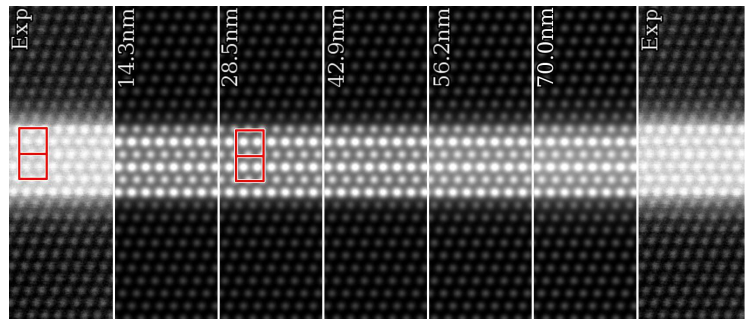}}
\hfill\

\hfill
\subfigure[Disordered \ce{AlAg2}\label{fig-simulation-disordered}]{\includegraphics[width=0.25\textwidth,angle=90]{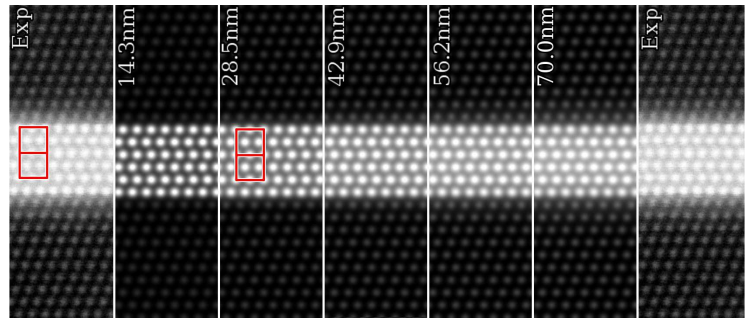}}
\hfill\

\hfill
\subfigure[Intensity profile for ordered \ce{AlAg2} \label{fig-profile-ordered}]{\includegraphics[width=11cm]{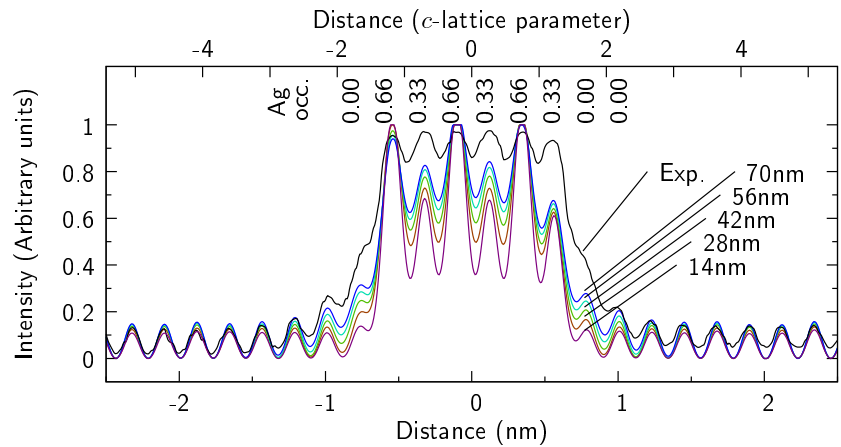}}
\hfill\

\hfill
\subfigure[Intensity profile for  disordered \ce{AlAg2}\label{fig-profile-disordered}]{\includegraphics[width=11cm]{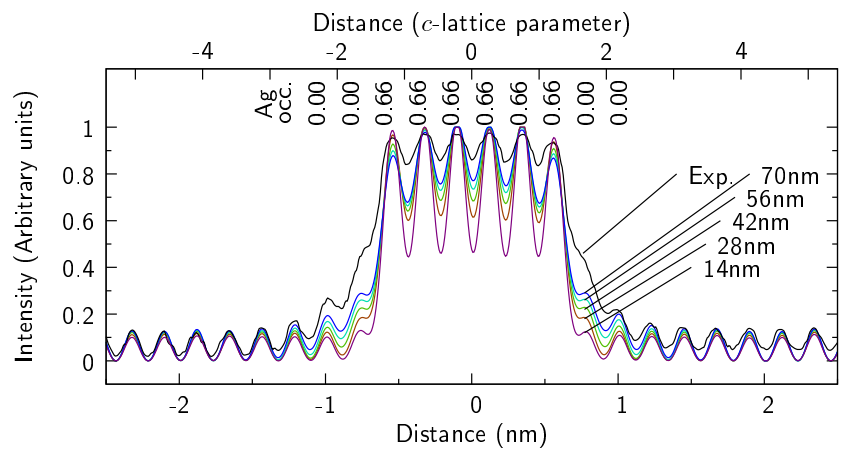}}
\hfill\

\caption{HAADF-STEM images and simulations for (a) ordered  and (b) disordered \gammaprime precipitates. 
The recorded image (labelled ``Exp'') is compared with simulations for  foil thicknesses of 14--70\,nm. The outline indicates a single hcp unit cell. 
Intensity profiles are shown in (c) and (d), respectively. The  simulated foil thickness and Ag occupancy is indicated.
\label{fig-simulation}}
\end{center}
\end{figure}

Simulated HAADF-STEM images of a  2c(\gammaprime) precipitate plate embedded in an aluminium matrix are provided in Figure \ref{fig-simulation}.
The figure compares an experimentally-obtained image with a series of multi-slice simulations for foil thicknesses of 14--70\,nm.  
The maximum foil thickness was selected based on position-averaged CBED patterns obtained in STEM mode. A comparison between these images and simulated  CBED patterns indicates thicknesses of $\le70$\,nm.
The precipitate is assumed to have the full thickness of the foil.
Simulations calculated with the Neumann model (Figure~\ref{fig-simulation-ordered}) predict strong differences in contrast between alternate basal planes for foil thickness, which do not appear in the experimentally-obtained micrograph.
Simulations calculated with a disordered structure (Figure~\ref{fig-simulation-disordered}) produce images that more closely match the experimental image.

Line profiles of HAADF-STEM intensity normal to the habit plane for ordered and disordered structures are provided in Figures~\ref{fig-profile-ordered} and \ref{fig-profile-disordered}.
After linear scaling and gamma-correction, the simulations reproduce the peak intensity for Al and high Ag layers with good accuracy, although the intensity in the troughs between Ag columns is greater in the experimental image, possibly due to detector noise.
The simulated foil thickness and Ag occupancy is indicated for each profile and it is clear that low Ag (occupancy of 0.33) layers in Figures~\ref{fig-profile-ordered} should be readily distinguishable, suggesting that long-range order should be clearly resolvable in the present instrumentation. On this basis, the HAADF-STEM images indicate that the  \gammaprime precipitates do not have long-range ordering of Ag on alternating basal planes.

Intensity profiles for the disordered structure (Figures~\ref{fig-profile-disordered}) compare well with the experimental image, except at the matrix layers immediately adjacent to the precipitate, where the measured intensity is well in excess of that expected from the simulations. 
Additional simulations examine whether this additional HAADF-STEM intensity at the interface is due to Ag segregation to the \gammaprime precipitate. 
Two models for segregation are considered, using the disordered model in Figure~\ref{fig-simulation-disordered} as a basis.
In the first, ``monolayer'' model the  matrix  layer adjacent to the precipitate is given an Ag occupancy of 0.333, half of that in the precipitate. 
The second, ``bilayer'' model mimics more diffuse segregation. The fcc layers adjacent to the precipitate is given an Ag occupancy of 0.33, and the following layer has an Ag occupancy of 0.167.

\begin{figure}[htbp]
\begin{center}
\hfill
\begin{minipage}[c]{0.48\textwidth}
\subfigure[Monolayer Ag segregation \label{fig-simulation-segregation1}]{\includegraphics[width=0.4\textwidth]{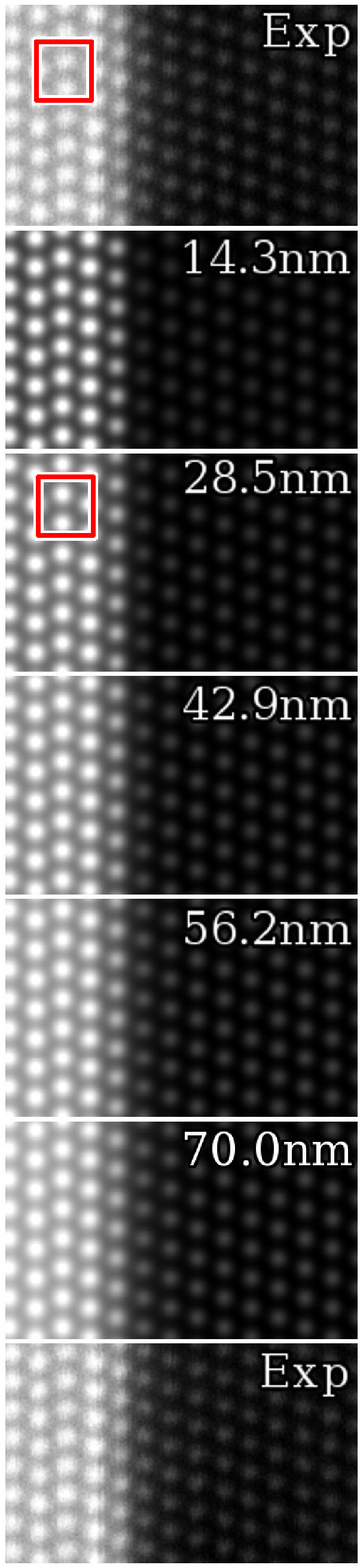}}
\hfill
\subfigure[Bilayer Ag segregation\label{fig-simulation-segregation2}]{\includegraphics[width=0.4\textwidth]{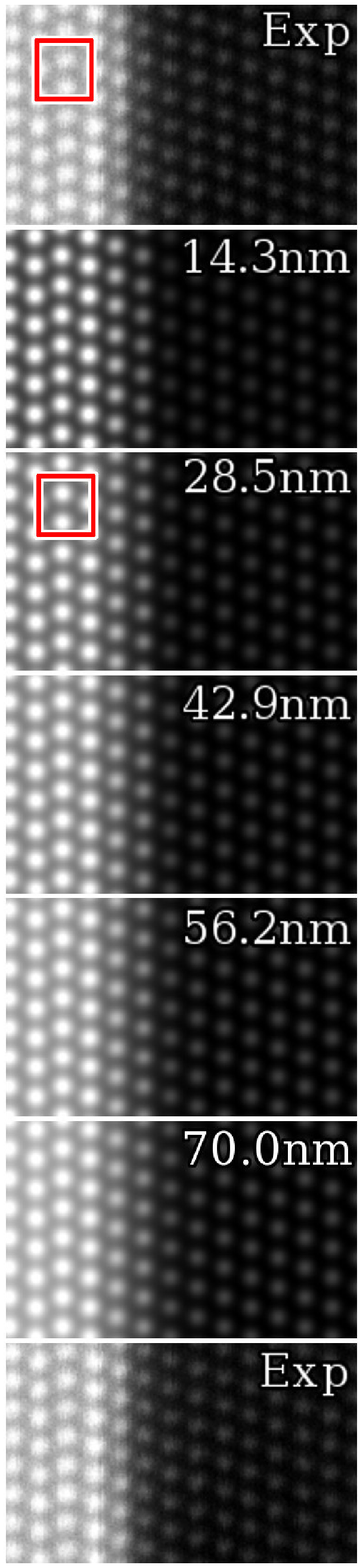}}
\hfill
\end{minipage}~
\begin{minipage}[c]{0.48\textwidth}
\subfigure[Intensity profile for  monolayer Ag segregation  \label{fig-profile-segregation1}]{\includegraphics[width=0.9\textwidth]{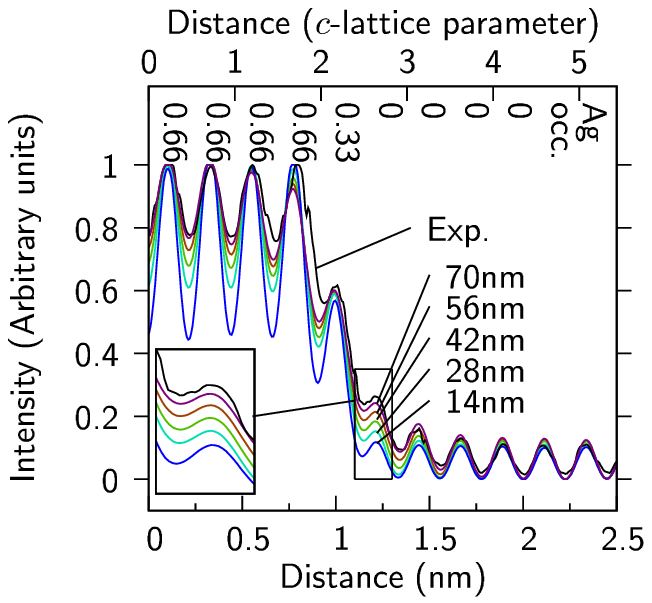}}

\subfigure[Intensity profile for  bilayer Ag segregation \label{fig-profile-segregation2}]{\includegraphics[width=0.9\textwidth]{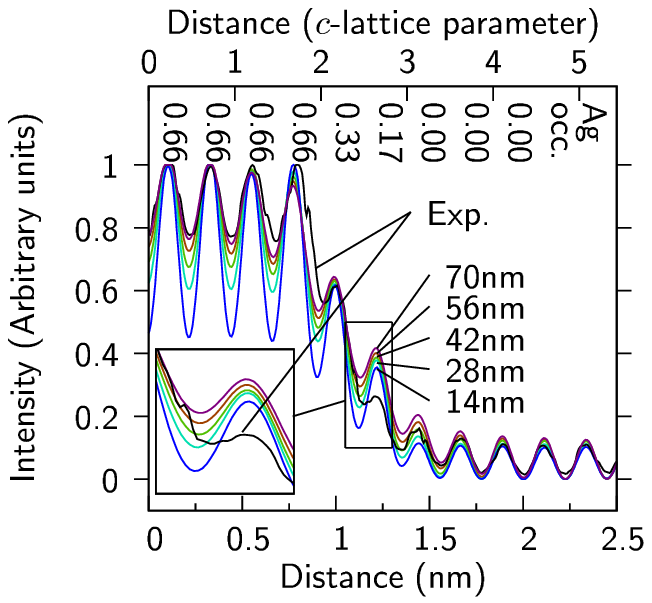}}

\end{minipage}
\caption{HAADF-STEM images and simulations of Ag segregation to a 3c(\gammaprime) precipitate.
(a) and (b) show HAADF-STEM images and simulations of Ag segregation to the matrix-precipitate interface.
A micrograph of a 3c(\gammaprime) precipitate (``Exp'') is compared with simulations for  (a)  monolayer and (b) bilayer Ag segregation to the interface. The outline indicates a single hcp unit cell. 
Intensity profiles are shown in (c) and (d), respectively. The  simulated foil thickness and Ag occupancy is indicated.
\label{fig-simulation-1}}
\end{center}
\end{figure}

The simulations including the effect Ag segregation to the precipitate are shown in Figure~\ref{fig-simulation-1}.
The figure compares an experimentally recorded image of the  precipitate-matrix interface of a 3c(\gammaprime) precipitate (Fig~\ref{1c}) with simulations using monolayer and bilayer segregation.
Simulations with monolayer segregation are shown in Figure~\ref{fig-simulation-segregation1}.
The corresponding intensity profiles provided in Figure~\ref{fig-profile-segregation1} show that this model for Ag segregation compares well with the experimental data. 
An enlarged inset shows the  region corresponding to the fcc layer adjacent to the precipitate. 
The experimental profile is close to the simulated profile for a foil thickness of 70\,nm. 
Simulations with bilayer segregation  (Figure~\ref{fig-simulation-segregation2}) have a broader, more diffuse interface than in the experimental image. 
The intensity profile in  Figure~\ref{fig-profile-segregation2}) shows this as greater HAADF-STEM intensity  in the second fcc layer adjacent to the precipitate. This is shown in greater detail in the enlarged inset which shows significantly greater HAADF-STEM intensity for simulations of all thicknesses than for the experimentally recorded micrographs.
 This implies that Ag segregates to the \gammaprime precipitates, but that the segregation is essentially limited to a monolayer adjacent to the precipitate.

\thispagestyle{empty}

\begin{figure}[htbp]
\begin{center}
\hfill
\subfigure[Weak segregation\label{fig-twincell-compare1}]{\includegraphics[width=0.3\textwidth]{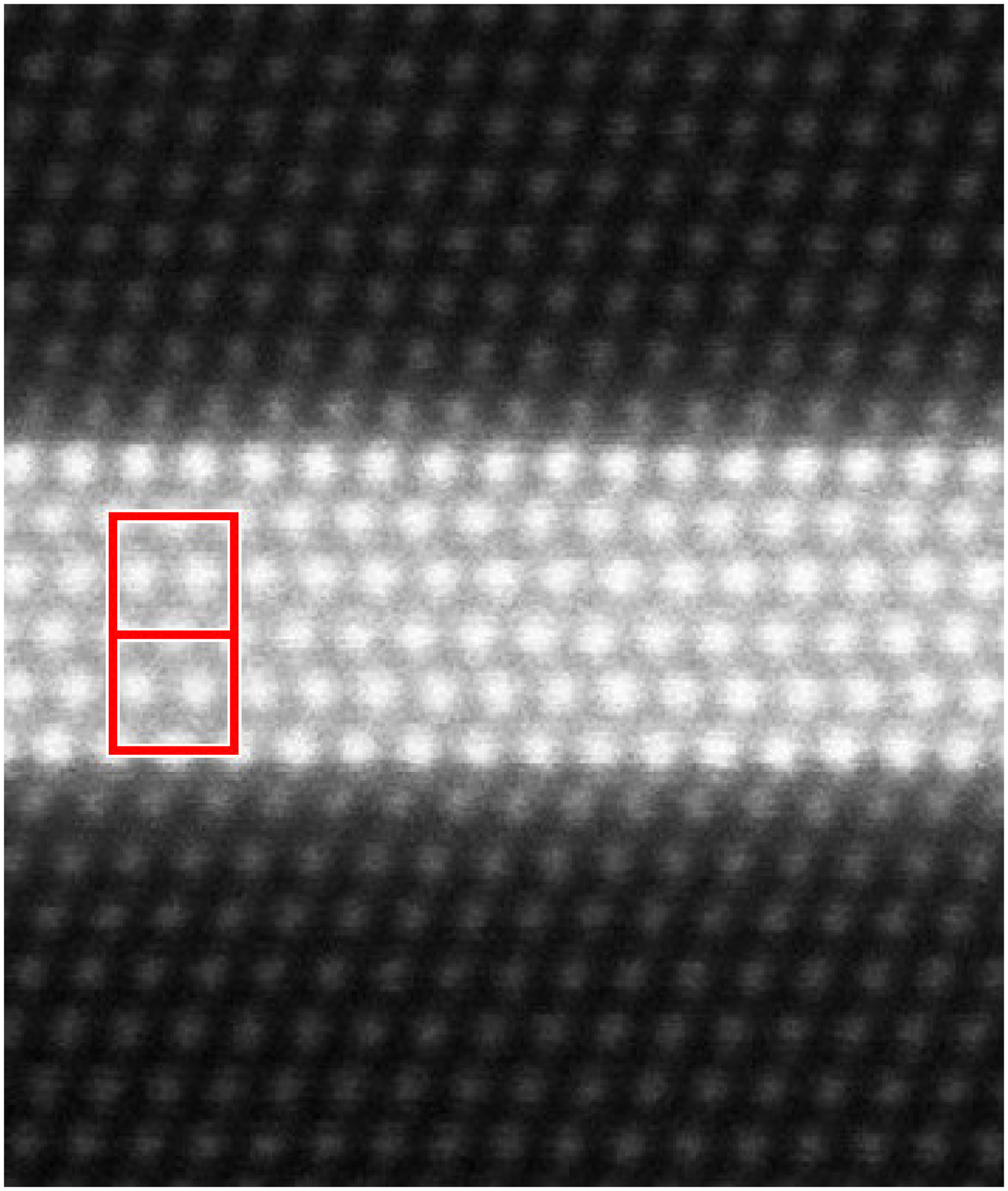}}
\hfill\
\subfigure[Strong segregation\label{fig-twincell-compare2}]{\includegraphics[width=0.29\textwidth]{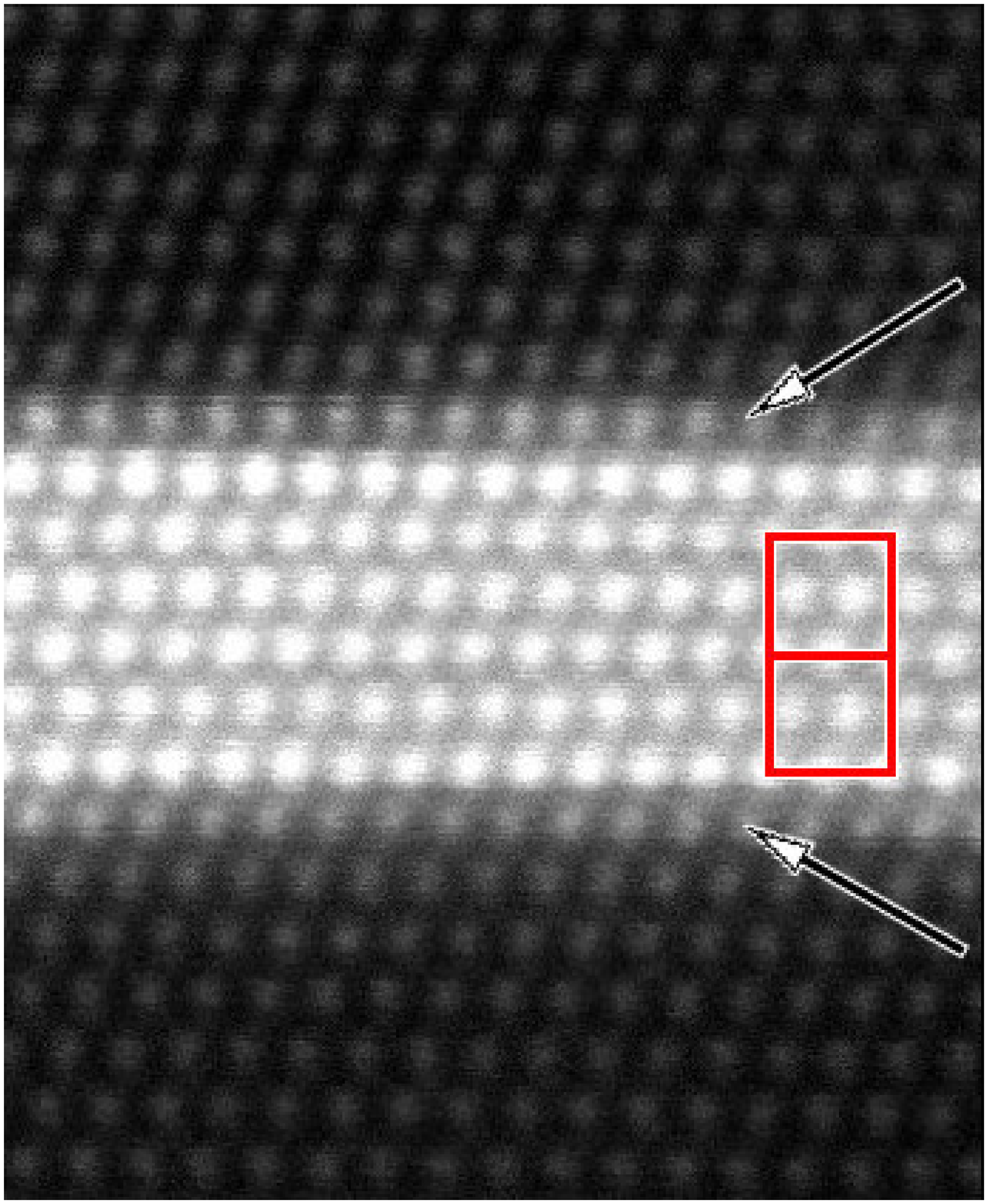}}
\hfill\
\subfigure[Intensity profiles\label{fig-twincell-compare3}]{\includegraphics[width=0.3\textwidth]{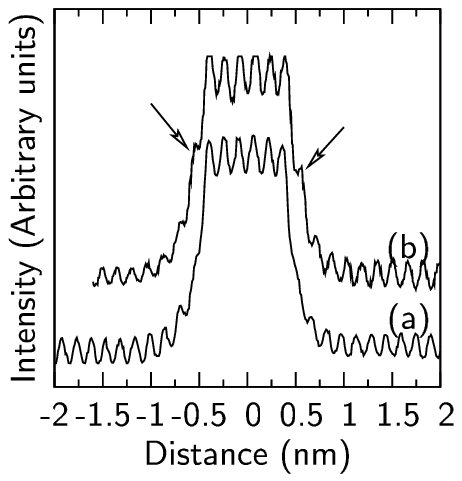}}
\hfill\
\caption{Ag segregation to 2c(\gammaprime) precipitates.
(a) and (b) show micrographs in which there is weak and strong Ag segregation to the precipitate, respectively.
Line profiles of the HAADF-STEM intensity show a strong shoulder and additional peaks in (b). 
\label{fig-twincell-compare}}
\end{center}
\end{figure}

The apparent thicknesses of precipitates varied considerably between precipitates having the same number of fcc$\rightarrow$hcp stacking faults.
Figure~\ref{fig-twincell-compare} compares two precipitates, each of which has two fcc$\rightarrow$hcp stacking faults, but with different apparent thicknesses.
The precipitate in Figure~\ref{fig-twincell-compare1} shows strong and near-equal intensity over six close-packed planes, with a slight increase in intensity in the adjacent fcc layer compared to the bulk matrix. 
Figure~\ref{fig-twincell-compare2}  shows a precipitate with much more stronger intensity in these adjacent fcc layers (indicated by arrows).
Intensity profiles for the two precipitates are provided in Figure~\ref{fig-twincell-compare3}  and show the additional intensity in (b) as distinct peaks of high, but unequal intensity. 
This indicates that the extent of Ag segregation to the \gammaprime phase can differ between precipitates, possibly due to local availability of solute.

\begin{figure}[htbp]
	\begin{center}
		\includegraphics{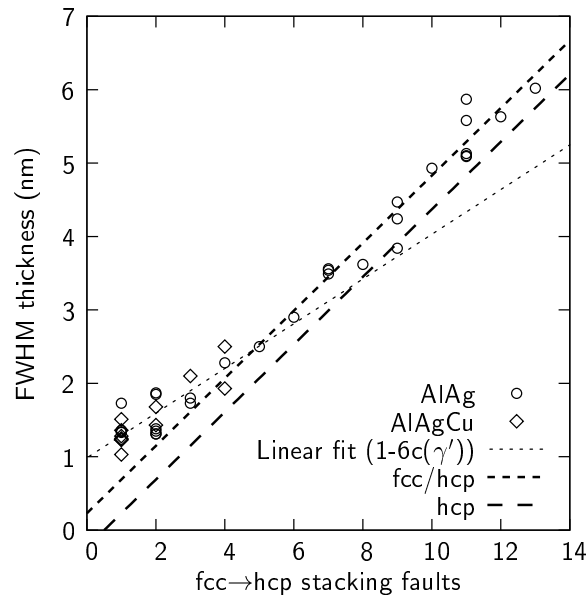}
		\caption{The full width half maximum (FWHM) thickness of the \gammaprime precipitate (as measured from the HAADF-STEM intensity) versus the number of stacking faults. Dashed lines indicate the width of the region having hcp stacking and the width including the fcc/hcp interface layers.
\label{fig-fwhm}}
	\end{center}
\end{figure}

The extent of Ag segregation to the precipitates was also measured by plotting the HAADF-STEM intensity normal to the coherent interface and comparing this to the width of the hcp region. 
The apparent thicknesses of the precipitates (defined as the full-width half-maximum (FWHM) of the high intensity region in HAADF-STEM micrographs) is plotted against the number of  stacking faults  in Figure~\ref{fig-fwhm}. 
Dashed lines in the figure indicate the width of the region with hcp stacking for a given number of stacking faults. 
For $n$ fcc$\rightarrow$hcp stacking faults there are 2$n$ hcp layers. The width of this hcp region is therefore
$(2n -1)$ times the layer spacing of 0.23\,nm. 
If the coherent interface (``fcc/hcp'') layers  on either side of the hcp region are included, the maximum width is $(2n+1)$ times the layer spacing.

The FWHM thickness of precipitates with 1--3 stacking faults is substantially greater than the width of the hcp region, even including the two common fcc/hcp  layers. 
This is most apparent for single faulted precipitates, where the average experimental FWHM value is  greater by approximately 0.5\,nm. 
The thin dashed line on the plot shows a linear fit to the data for precipitate with $\le$6 stacking faults which converges with the width of the hcp region at  approximately 5$c(\gammaprime)$.
Precipitates with  $\ge$7$c(\gammaprime)$ have FWHM thickness that generally lie close to the width of the hcp region.

Analysis of the HAADF-STEM profiles therefore indicates variable but quite substantial Ag occupancy in one fcc layer at the  planar \gammaprime-Al interface, with the segregation gradually diminishing in intensity as the precipitates thicken. 

\subsection{Convergent beam electron diffraction}

\begin{figure}[htbp]
	\begin{center}
		\hfill
		\subfigure[Sample region \label{fig-cbed-series-a}]{\includegraphics[width=0.35\textwidth]{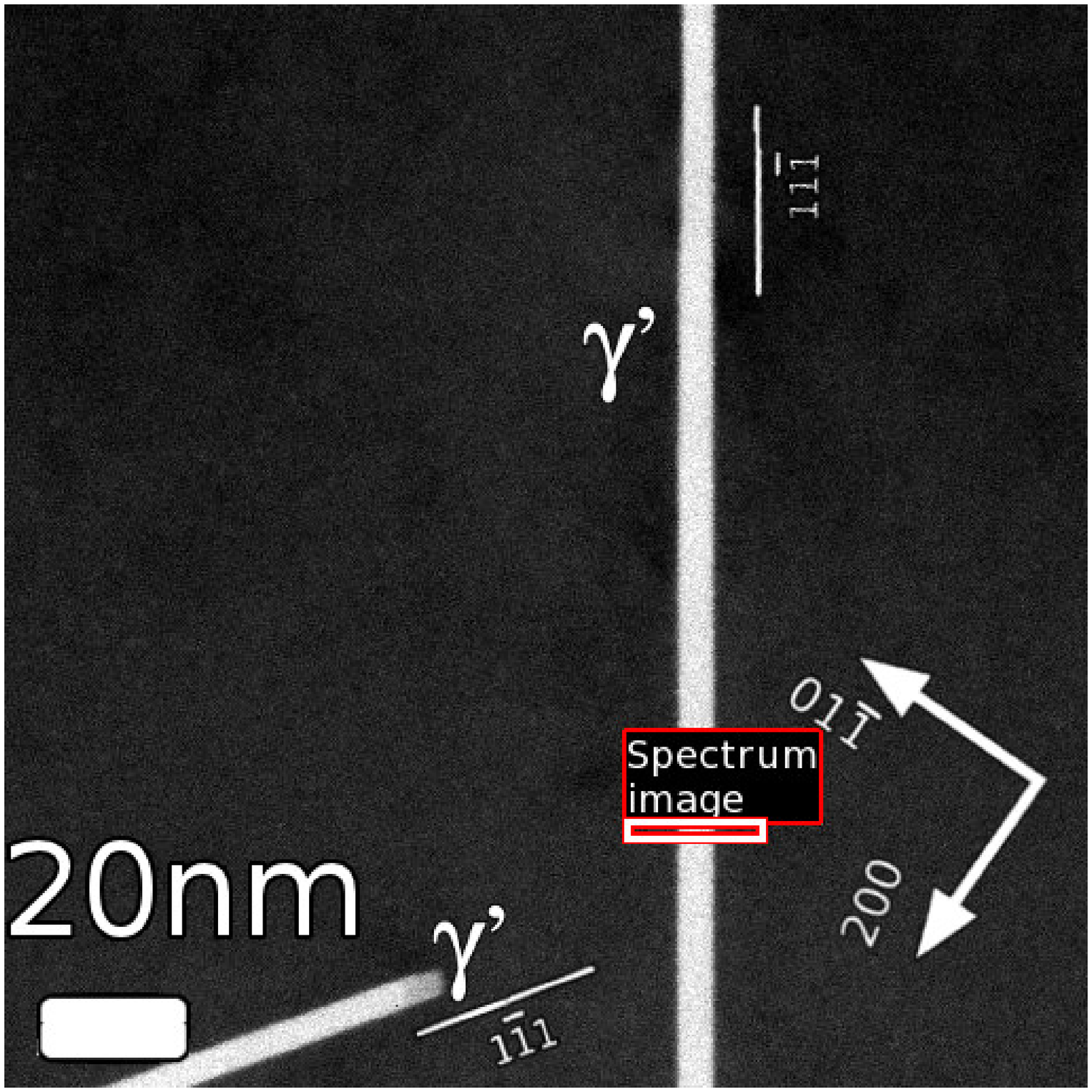}}
		\hfill
		\subfigure[Matrix CBED \label{fig-cbed-series-b}]{\includegraphics[width=0.35\textwidth]{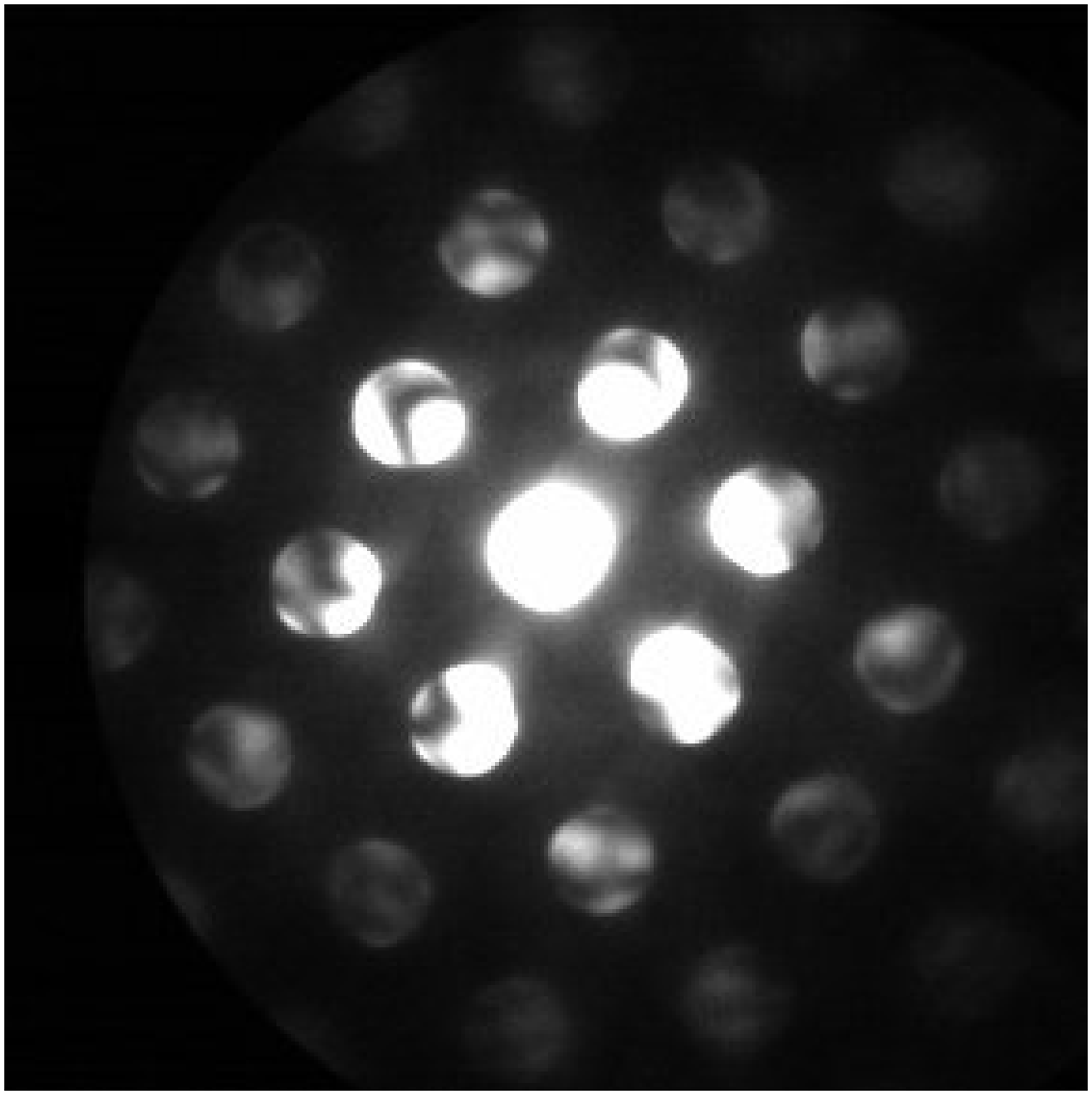}}
		\hfill\

		\hfill
		\subfigure[Precipitate CBED\label{fig-cbed-series-c}]{\includegraphics[width=0.35\textwidth]{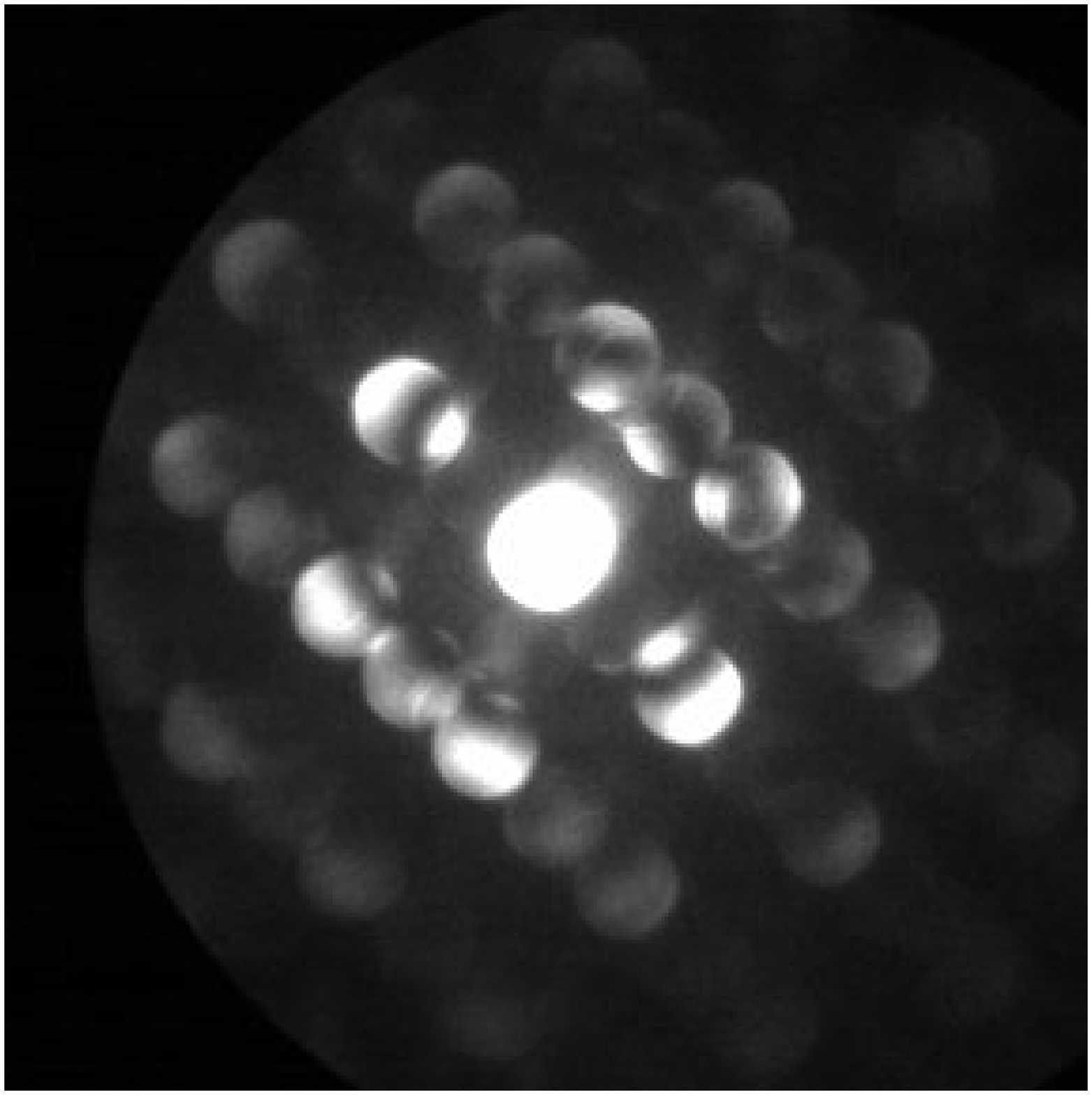}}
		\hfill
		\subfigure[Assignment of diffraction discs \label{fig-cbed-series-d}]{\includegraphics[width=0.35\textwidth]{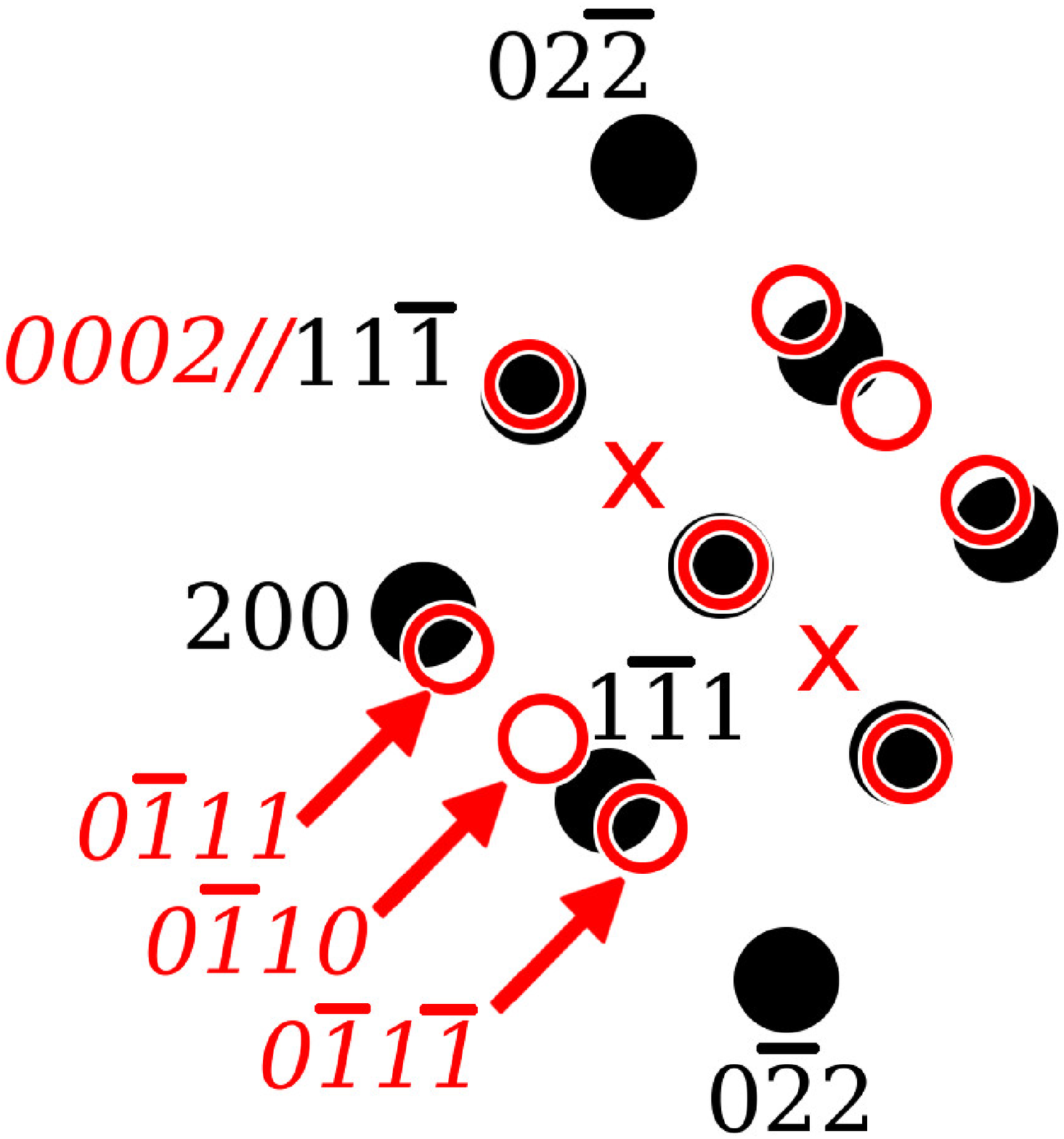}}
		\hfill\

		\caption{Convergent beam electron diffraction pattern of a \gammaprime precipitate in a sample aged for 23\,h.  (a) shows a HAADF-STEM image of the region of interest, with the \gammaprime precipitate running from top to bottom of the micrograph.  
(b) shows the CBED pattern for the matrix adjacent to the precipitate.
(c) shows a CBED pattern from the \gammaprime precipitate.
The main diffracted discs are assigned in (d). 
Filled and open circles indicate matrix and precipitate reflections, respectively.
Note the absence of diffracted intensity in the $\pm0001(\gammaprime)$ positions (labelled ``x'' in (d)). 
\label{fig-cbed-series}}
	\end{center}
\end{figure}

Convergent beam electron diffraction patterns show no evidence of chemical order of \gammaprime precipitates aged for 23\,h at 473\,K.
Maps of CBED patterns were obtained across a 20\,nm region containing a \gammaprime precipitate with a thickness of approximately 5\,nm, as shown in the STEM image in Figure~\ref{fig-cbed-series-a}.
The matrix and precipitate CBED patterns shown in Figure~\ref{fig-cbed-series-b} and (c) respectively are position-averaged over a small area within the matrix and precipitates. 
Figure~\ref{fig-cbed-series-d} indicates the location and indices of the diffraction disks observed on the CBED patterns. 
They are consistent with the orientation relationship: $(11\overline{1})// (0002);[011]//[2\overline{11}0]$. 
The precipitate CBED pattern in Figure~\ref{fig-cbed-series-c} shows no significant diffracted intensity for (0001)$_{\gammaprime}$ reflections, the presence of which would be a strong indication of compositional variations between adjacent basal planes.
Therefore the CBED observations are consistent with the HAADF-STEM imaging and simulations in revealing a lack of chemical long-range order within the precipitates. 

\section{Discussion}

\subsection{Solute segregation}

The micrographs shown in this work highlight the difficulty in unambiguously defining the thickness of the \gammaprime precipitates. This arises from the similarity between the fcc and hcp structures (which are interchangeable via the introduction of stacking faults) and the coherence of the matrix-precipitate interface. 

Throughout this work, the thickness of the \gammaprime precipitates is deliberately defined by the number of changes in stacking sequence required to produce the local structure from a pure fcc matrix. 
This structure-based definition allows the thickness of the Ag-enriched region to be compared with the local stacking sequence.
Figure~\ref{fig-stacking-summary} illustrates this schematically for precipitates with 1, 2, 3 and $n$ stacking faults. 

Density functional theory calculations of Ag close to a stacking fault in Al \cite{finkenstadt:2006} showed that the lowest energy sites for Ag were those with hcp stacking (i.e. the adjacent layers have identical stacking, for example a ``B'' layer sandwiched between two ``A'' layers). 
A single fcc$\rightarrow$hcp stacking fault (Figure~\ref{twin-1}) results in the formation of two hcp layers.
The addition of a further stacking fault in (Figure~\ref{twin-2})  generates an additional 2 close-packed layers with hcp structure and an arbitrary number of such faults there will generate $2n$ close-packed layers with hcp structure (Figure~\ref{twin-n}).
Regardless of the size of the precipitate the two surrounding layers are common to fcc and hcp structures; however these should not be assumed to be energetically equivalent to the hcp layers. 

The existence of both  hcp layers and interface fcc/hcp layers makes it difficult to clearly define the thickness of the precipitate based solely on the width of the Ag-enriched region. 
For example, for a single stacking fault (Figure~\ref{twin-1}) there are 2  hcp layers and 2 interface fcc/hcp layers. 
This would seem to permit a \gammaprime precipitate of thickness of between 2 and 4 layers (i.e. 0.5 and 1.5 times the $c$-lattice parameter).
However, HAADF-STEM images of single stacking-fault (1c(\gammaprime)) precipitates (e.g. Figure~\ref{1c}) show strong intensity over 5--6 atomic layers,  indicating that there is also substantial Ag in fcc atomic layers.
The width of this Ag-enriched zone cannot, therefore, be explained by the presence of the \gammaprime precipitate alone.
The image simulations in Figure~\ref{fig-simulation}  indicate that this does not result from intensity transferred from the precipitate atomic columns. 
The number of atomic layers exhibiting stronger intensity can be reproduced by simulations which include partial Ag occupancy in the close-packed fcc layers at the interface (Figure~\ref{fig-simulation-1}) and hence it is taken as good evidence of Ag segregating to the broad coherent interfaces of the precipitate.
This effect is most clear for precipitates with 1--3 stacking faults and appears to diminish with increasing precipitate thickness (See Figure~\ref{fig-fwhm}). 

\begin{figure}[bhtp]
\hfill
\subfigure[One stacking fault/1.5c \label{twin-1}]{
	\begin{tabular}{ccc}
		\\
		\\
		Structure & Stacking  \\ \toprule
	        fcc  		&		C 	 		\\
		fcc/hcp	&   A 				\\
		hcp		& \textbf{B}		 \\ \cmidrule(rl){2-2}
		hcp		& \textbf{A}		 \\
		fcc/hcp	&   B 				\\
	        fcc  		&		C 	 		\\
	\bottomrule
	\end{tabular}
}\hfill
\subfigure[Two stacking faults/2.5c \label{twin-2}]{
	\begin{tabular}{cc}
		Structure & Stacking \\ \toprule
	        fcc  		&		C 	 		\\
		fcc/hcp	&   A 				\\
		hcp		& \textbf{B}		 \\ \cmidrule(rl){2-2}
		hcp		& \textbf{A}		 \\
		hcp		& \textbf{B}		 \\ \cmidrule(rl){2-2}
		hcp		& \textbf{A}		 \\
		fcc/hcp	&   B 				\\
	        fcc  		&		C 	 		\\
	\bottomrule
	\end{tabular}
}\hfill\

\hfill
\subfigure[Three~stacking~faults/3.5c \label{twin-3}]{
	\begin{tabular}{cc}
		Structure & Stacking  \\ \toprule
	        fcc  		&		C 	 		\\
		fcc/hcp	&   A 				\\
		hcp		& \textbf{B}		 \\ \cmidrule(rl){2-2}
		hcp		& \textbf{A}		 \\
		hcp		& \textbf{B}		 \\ \cmidrule(rl){2-2}
		hcp		& \textbf{A}		 \\
		hcp		& \textbf{B}		 \\ \cmidrule(rl){2-2}
		hcp		& \textbf{A}		 \\
		fcc/hcp	&   B 				\\
	        fcc  		&		C 	 		\\
	\bottomrule
	\end{tabular}
}\hfill
\subfigure[$n$ stacking faults/$n$+0.5c\label{twin-n}]{
	\begin{tabular}{cc}
		\\
		\\
		\\
		\\
		Structure & Stacking  \\ \toprule
	        fcc  		&		C 	 		\\
		fcc/hcp	&   A 			\\
		hcp		& \textbf{B}		 \\
		\dots & 	 \dots   		  \\
		hcp		& \textbf{A}		 \\
		fcc/hcp	&   B 				\\
	        fcc  		&		C 	 		\\
	\bottomrule
	\end{tabular}
}\hfill\

\caption{Stacking sequences and local structures for a hcp precipitate embedded in a fcc matrix. 
The introduction of $n$ fcc$\rightarrow$ hcp stacking faults generates $2n$ hcp layers (indicated in bold type). The surrounding interface layer is common to the fcc and hcp structures. 
\label{fig-stacking-summary}}
\end{figure}

\subsection{Long-range order}

There is no measurable long-range chemical order in alternate basal planes of \gammaprime precipitates in alloys isothermally aged at 473\,K. 
Multi-slice image simulations indicate that HAADF-STEM is highly sensitive to local composition and should be readily capable of detecting long-range chemical order in  Neumann-structure \gammaprime precipitates for foil thickness $\ge$ 15\,nm.
Simulated intensity profiles for the ordered Neumann structure (Figure~\ref{fig-simulation}) are sensitive to layer composition and the intensity alternates between high and low values. 
Despite this, in experimentally obtained HAADF-STEM profiles,  Ag-rich columns all have relative intensities of  0.94--1.0, with 
only slight fluctuations in the intensity of adjacent basal planes  and no systematic intensity changes. 
Together with the lack of diffracted intensity for the $\pm$0001 diffracted discs  in CBED this is strong evidence that the structure lacks long-range order. This is consistent with studies using X-ray radiation which reported compositional differences of $\le$1.5\%Al between adjacent basal planes \cite{neumann:1966}.

It has been proposed that the appearance of alternating contrast in HRTEM studies \cite{howe:1987a} could arise from  localised strains around the precipitates \cite{YuGammaprime2004}.
However, it should be noted that the high-resolution studies reporting alternating strong and weak contrast on basal planes were conducted on alloys aged at 623\,K.
Further investigation on \gammaprime precipitates aged at this temperature would be required to determine whether the reported  ordering might arise from a kinetic effect at higher temperatures.

\subsection{Stacking faults}

Genuine hcp$\rightarrow$fcc stacking faults are rarely observed in \gammaprime precipitates and are seen only in thicker precipitates, such as the example shown in Figure~\ref{fig-stacking-3}. 
The rarity of such faults is in contrasts to the high fault densities reported previously \cite{guinier:1942,Guinier1952,borchers:1969,NicholsonNutting1961}.

Earlier reports of stacking faults in \gammaprime precipitates can be explained by the presence of an Ag atmosphere around the hcp region of the \gammaprime precipitates, as noted in Figures~\ref{fig-haadf-series} and \ref{fig-stacking-2} and the simulations in Figures~\ref{fig-simulation} and \ref{fig-simulation-1}.
Classically, the solute level around a precipitate is less than in the bulk matrix (See, for example \cite{moore:2000}),  however, \textit{ab initio} studies indicate that Ag segregates preferentially to hcp sites \cite{finkenstadt:2006}. Furthermore,  solute segregation to \gammaprime precipitates was noted by Osamura \textit{et al.} who reported Ag levels of 0.14\,at.\% close to 1--2 layer \gammaprime precipitates, compared to a  concentration of 0.066\,at.\% Ag in the matrix \cite{Osamura1986}.

The Ag level in the interface region is not reported quantitatively here as this would require careful thickness measurements and STEM detector calibration. 
The non-stoichiometric composition of the precipitate  \cite{Osamura1986,moore:2000} would be a further complication. 
However, the HAADF-STEM intensity profiles suggest that for precipitates of thickness $\le3c$ the level of Ag adjacent to the hcp regions may be around half that of the precipitate.

The Ag-rich interface region could be considered as a stacking fault in the outer layers of the \gammaprime precipitate and would indeed appear to be so in diffraction. 
However, there is no boundary dislocation separating the Ag-rich fcc region from the matrix and it is more valid to regard these layers as Ag segregation to a purely hcp \gammaprime precipitate.

The separation of the Ag-rich interface layers is in agreement with the stacking-fault spacing reported in \gammaprime precipitates. 
In precipitates of thickness 1-2$c(\gammaprime)$, the Ag-rich fcc layers are separated by 5--6 atomic layers (0.92--1.15\,nm).
This is in good agreement with a fault spacing of  1\,nm  reported in  thin precipitates  \cite{guinier:1942,Guinier1952,borchers:1969}.
Moreover, since these layers are only present at the broad precipitate-matrix interfaces, their spacing will increase as the precipitate thickens, explaining Nicholson and Nuttings' observation that the average fault spacing increased with the precipitate thickness \cite{NicholsonNutting1961}.

It appears surprising that the Ag-rich interface layers have the fcc structure, especially since Ag has been shown to lower the local stacking fault energy in Al \cite{Schulthess1998}. 
However, it should be noted that the \gammaprime precipitates are likely to have non-stoichiometric compositions and that the hcp structure only becomes preferred for  50--90 at.\% Ag \cite{Schulthess1998}.  
If the precipitate (and surrounding Ag-rich interface layer) have compositions similar to previous analyses  \cite{Osamura1986,moore:2000} it would explain why the Ag-segregated interface layers retain the fcc structure.

\subsection{Growth mechanism and kinetics}

The results provide valuable insight into the growth mechanism and kinetics of \gammaprime phase precipitation. 
The growth of large \gammaprime precipitates in Al-22at.\%Ag has been shown to be diffusion-controlled with solute depletion at the broad interfaces \cite{moore:2000,Moore2002}. 
However, a dramatically different situation exists for  1--3$c(\gammaprime)$ thickness, which are surrounded by Ag concentrations possibly up to half the Ag level in the precipitate itself (See Figure~\ref{fig-simulation-1}). 
This clearly indicates that growth is not limited by the supply of solute and must instead be controlled by the rate at which the solute crosses the matrix-precipitate interface, that is, interfacially controlled growth.

The Al-\gammaprime interface migrates via the nucleation and propagation of ledges on the broad, planar faces of the precipitates \cite{howe:1985a, howe:1987,moore:2000,Sagoe-crentsil1991,AikinPlichta1990}. 
Earlier studies indicated ledge heights or 2, 4 or 6 unit cells \cite{howe:1985a,AikinPlichta1990}; 
  however  growth via single unit cell ledges has also been proposed  based on the observation of  \gammaprime precipitates with thickness of 1, 2, and 3$c(\gammaprime)$  \cite{RosalieActa2011}.
The HAADF-STEM images and intensity profiles reported in this work (Figures~\ref{fig-haadf-series}--\ref{fig-line-profiles}) support the latter proposition, showing a series of \gammaprime precipitates with thickness increasing in integer multiples of the unit cell thickness up to 12$c(\gammaprime)$. 
Furthermore, single unit cell ledges are imaged for the first time (Figure~\ref{fig-haadf-series1}) and it seems clear that growth  initially occurs via single unit cell ledges. This will introduce an additional shear strain energy barrier to ledge nucleation when  the shear strains are not self-accommodating  (i.e. whenever the number of hcp$\rightarrow$fcc stacking faults is not a multiple of 3) \cite{RosalieActa2011}.

The high coherency and low interfacial energy \cite{Ramanujan1992a}  of the \gammaprime-Al interface make the nucleation of single unit cell ledges unfavourable and the scarcity of ledges in precipitates of 1-12$c(\gammaprime)$  suggests that ledge propagation is rapid compared to nucleation.  
It is probable, therefore, that interface migration during the early stages of \gammaprime phase growth is controlled by the nucleation rate of new ledges.

\section{Conclusions}
\begin{enumerate}
\item A semi-quantitative analysis of HAADF-STEM images showed no evidence of systematic long-range chemical order within \gammaprime precipitates isothermally aged at 473\,K for 2--23\,h. A comparison between image simulations and experimental images suggests that variations in composition between adjacent layers are non-systematic in nature and negligible in amount. 
The absence of (0001)$\gammaprime$ reflections in CBED also indicates that there is no measurable compositional difference between adjacent basal planes. 
\item For precipitates with 1--3 fcc$\rightarrow$hcp stacking faults the thickness of the Ag-enriched region is commonly considerably wider than the size of the hcp region. Image simulations indicate that this is not due to some of the scattered intensity spilling over from neighbouring atomic columns. Simulations with partial Ag occupancy in 1 fcc layer at the matrix-precipitate interface showed similar broader, diffuse interfaces and it is concluded that this indicates Ag segregation to the precipitates in the early stages of growth. This appears to be confined to one fcc layer on the broad coherent interfaces and generates an interface that is  structurally well-defined, but compositionally diffuse. 

\item In precipitates with $\ge$6 stacking faults the Ag-enriched zone region was approximately equal to the width of the hcp region, indicating that there was little solute segregation to these larger precipitates. 
\item The \gammaprime phase nucleates as single unit-cell thickness platelets (structurally equivalent to a single, Ag-enriched stacking fault) and thickens via the nucleation and growth of ledges with a height equal to one unit cell of \gammaprime. 
Ledges are uncommon, even on thicker precipitates, suggesting that ledge nucleation is slow relative to the lateral propagation of the ledges. 
\item Growth of \gammaprime precipitates via ledges of single unit cell riser height requires that precipitates progress through stages in which the transformation strain is not self-accommodated, which may contribute to the slow thickening rate of the precipitates.
\item hcp$\rightarrow$fcc stacking faults within the precipitates are rare. However, due to Ag segregation, the precipitates are surrounded by Ag-rich layers that retain the fcc structure. This is likely to be responsible for earlier reports of high densities of  stacking faults in \gammaprime precipitates.
\item The abundance of excess Ag at the precipitate-matrix interfaces implies that the growth of \gammaprime precipitates is not  initially controlled by the availability of solute (i.e. diffusional control) but is instead subject to interface control, most likely due to the difficulty of nucleating additional ledges.
\end{enumerate}

\section*{Acknowledgements}
The authors gratefully acknowledge the support of the Australian Research Council through the Centre of Excellence for Design in Light Metals.
The authors acknowledge use of the facilities at the Monash Centre for Electron Microscopy and engineering support by Russell King.
We are also grateful to one of the reviewers, for several insightful comments on the manuscript.  
Finally we are thankful for the support and encouragement of Professor Barrington C. Muddle, whose work initiated this project.


\begin{thebibliography}{10}
\expandafter\ifx\csname url\endcsname\relax
  \def\url#1{\texttt{#1}}\fi
\expandafter\ifx\csname urlprefix\endcsname\relax\def\urlprefix{URL }\fi
\expandafter\ifx\csname href\endcsname\relax
  \def\href#1#2{#2} \def\path#1{#1}\fi

\bibitem{passoja:1971}
D.~E. Passoja, G.~S. Ansell, Acta Metall. 19~(11) (1971) 1253--1261.
\newblock \href {http://dx.doi.org/10.1016/0001-6160(71)90059-9}
  {\path{doi:10.1016/0001-6160(71)90059-9}}.

\bibitem{finkenstadt:2006}
D.~Finkenstadt, D.~D. Johnson, Phys. Rev. B 73~(2) (2006) 024101.
\newblock \href {http://dx.doi.org/10.1103/PhysRevB.73.024101}
  {\path{doi:10.1103/PhysRevB.73.024101}}.

\bibitem{RosalieActa2011}
J.~M. Rosalie, L.~Bourgeois, B.~C. Muddle, Acta Mater. 59~(19) (2011)
  7168--7176.
\newblock \href {http://dx.doi.org/10.1016/j.actamat.2011.08.001}
  {\path{doi:10.1016/j.actamat.2011.08.001}}.

\bibitem{voss:1999}
H.~Voss, G.~Schmitz, F.~Haider, Philos. Mag. A 79~(2) (1999) 423--436.

\bibitem{Finkenstadt2009}
D.~Finkenstadt, D.~D. Johnson, Mater. Sci. Eng. A 525~(1-2) (2009) 174--180.
\newblock \href {http://dx.doi.org/10.1016/j.msea.2009.07.004}
  {\path{doi:10.1016/j.msea.2009.07.004}}.

\bibitem{howe:1985a}
J.~M. Howe, H.~I. Aaronson, R.~Gronsky, Acta Metall. 33~(4) (1985) 649--658.

\bibitem{howe:1987}
J.~M. Howe, U.~Dahmen, R.~Gronsky, Philos. Mag. 56~(1) (1987) 31--61.

\bibitem{moore:2000}
K.~T. Moore, J.~M. Howe, Acta Mater. 48~(16) (2000) 4083--4098.
\newblock \href {http://dx.doi.org/10.1016/S1359-6454(00)00211-1}
  {\path{doi:10.1016/S1359-6454(00)00211-1}}.

\bibitem{Sagoe-crentsil1991}
K.~K. Sagoe-Crentsil, L.~C. Brown, Philos. Mag. A 64~(2) (1991) 429--441.

\bibitem{AikinPlichta1990}
R.~M. Aikin, Jr, M.~R. Plichta, Acta Metall. Mater 38~(1) (1990) 77 -- 93.
\newblock \href {http://dx.doi.org/10.1016/0956-7151(90)90136-5}
  {\path{doi:10.1016/0956-7151(90)90136-5}}.

\bibitem{howe:1987a}
J.~M. Howe, J. Metal 39~(9) (1987) 13--16.

\bibitem{sagoe-crentsil:1987}
K.~K. Sagoe-Crentsil, L.~C. Brown, Aluminium Alloys (1987) 634--636.

\bibitem{Ramanujan1992a}
R.~V. Ranamujan, J.~K. Lee, H.~I. Aaronson, Acta Metal. Mater 40~(12) (1992)
  3421--3432.

\bibitem{guinier:1942}
A.~Guinier, J Phys. Rad. 3 (1942) 124--136.

\bibitem{Guinier1952}
A.~Guinier, Z. Metall. 43~(6) (1952) 217--223.

\bibitem{NicholsonNutting1961}
R.~B. Nicholson, J.~Nutting, Acta Metall. 9~(4) (1961) 332 -- 343.
\newblock \href {http://dx.doi.org/10.1016/0001-6160(61)90227-9}
  {\path{doi:10.1016/0001-6160(61)90227-9}}.

\bibitem{borchers:1969}
H.~Borchers, G.~Thym, Z. Metall. 60~(4) (1969) 303--312.

\bibitem{neumann:1966}
J.~P. Neumann, Acta Metall. 14~(4) (1966) 505--511.
\newblock \href {http://dx.doi.org/10.1016/0001-6160(66)90318-X}
  {\path{doi:10.1016/0001-6160(66)90318-X}}.

\bibitem{zarkevich:2002}
N.~A. Zarkevich, D.~D. Johnson, A.~V. Smirnov, Acta Mater. 50~(9) (2002)
  2443--2459.

\bibitem{Osamura1986}
K.~Osamura, T.~Nakamura, A.~Kobayashi, T.~Hashizume, T.~Sakurai, Acta Metall.
  34~(8) (1986) 1563 -- 1570.
\newblock \href {http://dx.doi.org/10.1016/0001-6160(86)90101-X}
  {\path{doi:10.1016/0001-6160(86)90101-X}}.

\bibitem{YuGammaprime2004}
S.~Y. Yu, B.~Sch{\"o}nfeld, H.~Heinrich, G.~Kostorz, Prog. Mater. Sci 49~(3-4)
  (2004) 561--579.
\newblock \href {http://dx.doi.org/10.1016/j.pmatsci.2003.08.002}
  {\path{doi:10.1016/j.pmatsci.2003.08.002}}.

\bibitem{Finkenstadt2010}
D.~Finkenstadt, D.~D. Johnson, Phys. Rev. B 81~(1) (2010) 014113.
\newblock \href {http://dx.doi.org/10.1103/PhysRevB.81.014113}
  {\path{doi:10.1103/PhysRevB.81.014113}}.

\bibitem{DwyerStem2010}
C.~Dwyer, Ultramicroscopy 110~(3) (2010) 195--198.
\newblock \href {http://dx.doi.org/10.1016/j.ultramic.2009.11.009}
  {\path{doi:10.1016/j.ultramic.2009.11.009}}.

\bibitem{RosalieThetaPrimesilver2012}
J.~M. Rosalie, L.~Bourgeois, Acta Mater. 60~(17) (2012) 6033--6041.
\newblock \href {http://dx.doi.org/10.1016/j.actamat.2012.07.039}
  {\path{doi:10.1016/j.actamat.2012.07.039}}.

\bibitem{Schulthess1998}
T.~C. Schulthess, P.~E.~A. Turchi, A.~Gonis, T.-G. Nieh, Acta Mater. 46~(6)
  (1998) 2215--2221.
\newblock \href {http://dx.doi.org/10.1016/S1359-6454(97)00432-1}
  {\path{doi:10.1016/S1359-6454(97)00432-1}}.

\bibitem{Moore2002}
K.~T. Moore, W.~C. Johnson, J.~M. Howe, H.~I. Aaronson, D.~R. Veblen, Acta
  Mater. 50~(5) (2002) 943--956.
\newblock \href {http://dx.doi.org/10.1016/S1359-6454(01)00394-9}
  {\path{doi:10.1016/S1359-6454(01)00394-9}}.

\end{thebibliography}

\end{document}